\documentclass[aps,prl,twocolumn,superscriptaddress,amsmath,showpacs,tightenlines,pdflatex,longbibliography]{revtex4-1}
\usepackage{amssymb}
\usepackage{amsmath}
\usepackage{dcolumn}
\usepackage{graphicx}
\usepackage{mathrsfs}
\usepackage{subfigure}
\usepackage{booktabs}
\usepackage{color}

\usepackage{url}
\usepackage[colorlinks]{hyperref}
\hypersetup{%
    plainpages=true,
    breaklinks=true,
    hypertexnames=false,
    pageanchor=true,
    colorlinks=true,
    linkcolor={blue},
    citecolor={red},
    urlcolor={blue},
    anchorcolor={black}
}

\begin{document}

\title{Exponentially Enhanced Light-Matter Interaction, Cooperativities,
and Steady-State Entanglement Using Parametric Amplification}

\author{Wei Qin}
\affiliation{Quantum Physics and Quantum Information Division,
Beijing Computational Science Research Center, Beijing 100193,
China} \affiliation{CEMS, RIKEN, Wako-shi, Saitama 351-0198,
Japan}

\author{Adam Miranowicz}
\affiliation{CEMS, RIKEN, Wako-shi, Saitama 351-0198, Japan}
\affiliation{Faculty of Physics, Adam Mickiewicz University,
61-614 Pozna\'n, Poland}

\author{Peng-Bo Li}
\affiliation{CEMS, RIKEN, Wako-shi, Saitama 351-0198, Japan}
\affiliation{Shaanxi Province Key Laboratory of Quantum
Information and Quantum Optoelectronic Devices, Department of
Applied Physics, Xi'an Jiaotong University, Xi'an 710049, China}

\author{Xin-You L\"u}
\affiliation{School of Physics, Huazhong University of Science and
Technology, Wuhan 430074, China}

\author{J. Q. You}
\affiliation{Quantum Physics and Quantum Information Division,
Beijing Computational Science Research Center, Beijing 100193,
China}
\affiliation{Department of Physics, Zhejiang University, Hangzhou 310027, China}

\author{Franco Nori}
\affiliation{CEMS, RIKEN, Wako-shi, Saitama 351-0198, Japan}
\affiliation{Physics Department, The University of Michigan, Ann
Arbor, Michigan 48109-1040, USA}

\begin{abstract}
We propose an experimentally feasible method for enhancing the
atom-field coupling as well as the ratio between this coupling and
dissipation (i.e., cooperativity) in an optical cavity. It
exploits optical parametric amplification to exponentially enhance
the atom-cavity interaction and, hence, the cooperativity of the
system, with the squeezing-induced noise being completely
eliminated. Consequently, the atom-cavity system can be driven
from the weak-coupling regime to the strong-coupling regime for
modest squeezing parameters, and even can achieve an effective
cooperativity much larger than $100$. Based on this, we further
demonstrate the generation of steady-state nearly maximal quantum
entanglement. The resulting entanglement infidelity (which
quantifies the deviation of the actual state from a
maximally entangled state) is exponentially smaller than the lower
bound on the infidelities obtained in other dissipative
entanglement preparations without applying squeezing. In
principle, we can make an arbitrarily small infidelity. Our
generic method for enhancing atom-cavity interaction and
cooperativities can be implemented in a wide range of physical
systems, and it can provide diverse applications for quantum information
processing.
\end{abstract}
\maketitle
Cavity~\cite{Haroche2006Book} and
circuit~\cite{you2011atomic,gu2017microwave} quantum
electrodynamics (QED) provide promising platforms to implement
light-matter interactions at the single-particle level by
efficiently coupling single atoms to quantized cavity fields.
Exploiting such coupled systems for quantum information processing
often requires the strong-coupling regime (SCR), where the
atom-cavity coupling $g$ exceeds both atomic spontaneous-emission
rate $\gamma$ and cavity-decay rate $\kappa$, such that a single
excitation can be coherently exchanged between atom and cavity
before their coherence is lost. A typical parameter quantifying
this property is the cooperativity defined as
$C=g^{2}/\left(\kappa\gamma\right)$. Experimentally, microwave
systems (like quantum superconducting circuits) can have very high
cooperativities of order up to
$10^{4}$~\cite{gu2017microwave,schoelkopf2008wiring,xiang2013hybrid}.
However, for most optical systems (see~\cite{lev2004feasibility}
for a notable exception in photonic band gap cavities), it is
currently challenging to achieve the SCR and, in particular, the
cooperativity of $C$ larger than
$10^{2}$~\cite{park2006cavity,thompson2013coupling,shomroni2014all,
kato2015strong,hamsen2017two,welte2017cavity}. This directly
limits the ability to process quantum information in optical
cavities. Here, we propose a novel approach for this problem, and we
demonstrate that the light-matter coupling and cooperativity
can be exponentially increased with a cavity squeezing parameter.
Specifically, we parametrically squeeze the cavity mode to
strengthen the coherent coupling $g$, and at the same time, we
apply a broadband squeezed-vacuum field to completely eliminate
the noise induced by squeezing. As an intriguing application, we
show how to improve exponentially the quality of steady-state
entanglement.

Quantum entanglement is not only a striking feature of quantum
physics but also a fundamental resource in quantum information
technologies. The preparation of an entangled state between atoms
in optical cavities can be directly implemented using controlled
unitary
dynamics~\cite{pellizzari1995decoherence,zheng2000efficient}. However,
the presence of an atomic spontaneous emission and cavity loss leads
to a poor infidelity scaling
$\delta=\left(1-\mathcal{F}\right)\propto1/\sqrt{C}$~\cite{sorensen2003measurement},
where $\mathcal{F}$ is the fidelity, which characterizes the
distance between the ideal and actual states, and $\delta$ is the
corresponding infidelity. This is owing to the fact that both
decays can carry away information about the system and destroy its
coherence. For this reason, many approaches, which have been
proposed for entanglement preparation, are focused on dissipation
engineering, which treats dissipative processes as a resource
rather than as a detrimental
noise~\cite{bose1999proposal,chimczak2005teleportation,kraus2008preparation,verstraete2009quantum,krauter2011entanglement,burgarth2014exponential,yusipov2017localization,hartmann2017asymptotic}.
In the resulting entanglement, the infidelity scaling has a
quadratic improvement,
$\delta\propto1/C$~\cite{kastoryano2011dissipative,shen2011steady,lin2013dissipative,su2014scheme,reiter2016scalable,borregaard2015heralded,qin2017heralded}.
Such an infidelity, however, remains lower-bounded by the
cooperativity, because only partial dissipation contributes to the
entanglement, which still suffers errors from other kinds (or
channels) of dissipation. In this Letter, we demonstrate that
our approach for the cooperativity enhancement can lead to an
exponential suppression of undesired dissipation and, as a
consequence, of the entanglement infidelity. Since the discussed
model is generic, our proposal can be realized in a wide range of
physical systems, in particular, optical cavities.

\begin{figure}[tbph]
\centering
\includegraphics[width=8.4cm]{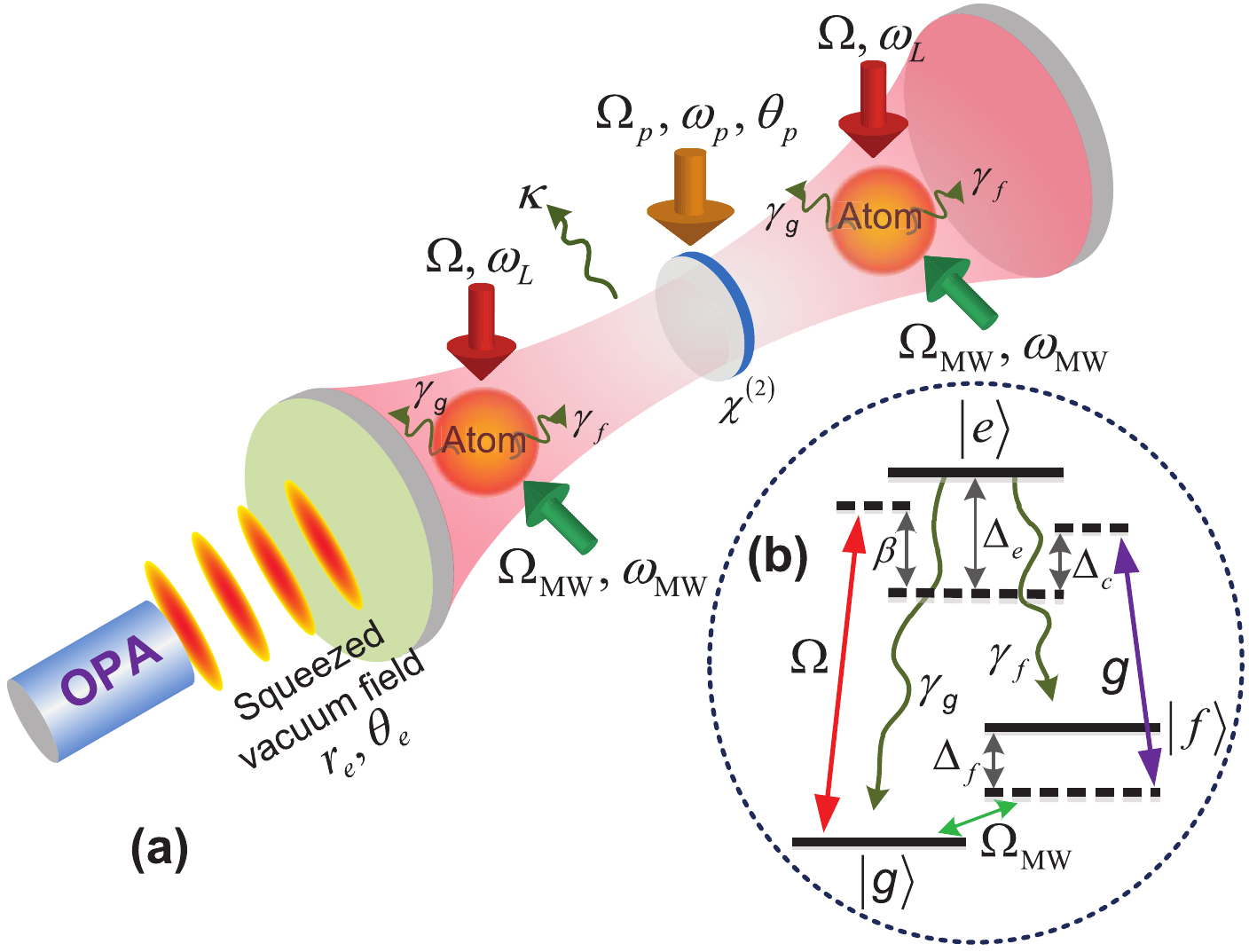}
\caption{Schematics of the proposed method for
enhancing cooperativity and  maximizing steady-state entanglement.
(a) Two driven atoms are trapped inside a single-mode cavity,
which contains a $\chi^{\left(2\right)}$ nonlinear medium strongly
pumped at amplitude $\Omega_{p}$, frequency $\omega_{p}$, and
phase $\theta_{p}$. The cavity couples to a squeezed-vacuum
reservoir, which is generated by optical parametric amplification
(OPA) with a squeezing parameter $r_{e}$ and a reference phase
$\theta_{e}$. As depicted in (b), the three-level atoms (in the
$\Lambda$ configuration) are coupled to the cavity mode with a
strength $g$. In addition, the transition with Rabi frequency
$\Omega$ ($\Omega_{\text{MW}}$) is driven by a laser (microwave)
field of frequency $\omega_{L}$ ($\omega_{\text{MW}}$). We also
assume that, along with a cavity decay rate $\kappa$, the excited
state $|e\rangle$ of the atoms decays to the ground states
$|g\rangle$ and $|f\rangle$ at rates $\gamma_{g}$ and
$\gamma_{f}$, respectively.}\label{figschematic}
\end{figure}

\emph{Basic idea.}---As depicted in Fig.~\ref{figschematic}(a),
we consider a quantum system consisting of two $\Lambda$ atoms and
a $\chi^{\left(2\right)}$ nonlinear medium. The atoms are confined
in a single-mode cavity of frequency $\omega_{c}$. The ground
states of each atom, $|g\rangle$ and $|f\rangle$, are excited to
the state $|e\rangle$, respectively, via a laser drive with Rabi
frequency $\Omega$ and the coupling to the cavity mode with
strength $g$, as shown in Fig.~\ref{figschematic}(b). If the
nonlinear medium is pumped (say, at frequency $\omega_{p}$,
amplitude $\Omega_{p}$, and phase $\theta_{p}$), then the cavity
mode can be squeezed along the axis rotated at the angle
$\left(\pi-\theta_{p}\right)/2$. When $\Omega_{p}$ is close to the
detuning $\Delta_{c}=\omega_{c}-\omega_{p}/2$, the atom-cavity
coupling can be enhanced exponentially with a controllable
squeezing parameter
$r_{p}=\left(1/4\right)\ln\left[\left(1+\alpha\right)/\left(1-\alpha\right)\right]$,
where $\alpha=\Omega_{p}/\Delta_{c}$. Meanwhile, squeezing the
cavity mode also induces thermal noise and two-photon correlations
in the cavity. In order to suppress them, a possible strategy is
to use the squeezed vacuum field to drive the cavity~\cite{murch2013reduction,bartkowiak2014quantum,lu2015squeezed,
lemonde2016enhanced, clark2017sideband,
zeytinouglu2017engineering}. This causes the squeezed-cavity mode to
equivalently interact with the thermal vacuum reservoir, and therefore, it
yields an effective cooperativity exhibiting an exponential
enhancement with $2r_{p}$.

Furthermore, to generate steady-state entanglement, we tune the
squeezed-cavity mode to resonantly drive the transition
$|f\rangle\rightarrow|e\rangle$, and as a result, the
excitation-number-nonconserving processes would be strongly
suppressed. Thus, in the limit of $\Omega\ll g_{s}$, the
ground-state subspace, spanned by
$\big\{|\phi_{\pm}\rangle=\left(|gg\rangle\pm|ff\rangle\right)|0\rangle_{s}/\sqrt{2},|\psi_{\pm}\rangle=\left(|gf\rangle\pm|fg\rangle\right)|0\rangle_{s}/\sqrt{2}\big\}$,
is decoupled from all of the excited states except the dark state,
$|D\rangle=\left(|fe\rangle-|ef\rangle\right)|0\rangle_{s}/\sqrt{2}$,
from the atom-cavity interaction. Here, the number refers to the
squeezed-cavity photon number. For entanglement preparation, in
order to be independent of an initial state, we apply an
off-resonant microwave field of frequency $\omega_{\text{MW}}$ to
couple $|g\rangle$ and $|f\rangle$ with the Rabi frequency
$\Omega_{\text{MW}}$, as shown in Fig.~\ref{figschematic}(b),
to drive the transitions
$|\phi_{-}\rangle\rightarrow|\phi_{+}\rangle\rightarrow|\psi_{+}\rangle$.
Subsequently, the laser drive $\Omega$ excites $|\psi_{+}\rangle$
to $|D\rangle$, which then decays to $|\psi_{-}\rangle$ via
atomic spontaneous emission. The populations initially in the
ground-state subspace are, thus, driven to and trapped in
$|\psi_{-}\rangle$, resulting in a maximally-entangled steady
state, the singlet state
$|S\rangle=\left(|gf\rangle-|fg\rangle\right)/\sqrt{2}$, between
the atoms. In contrast to previous proposals of entanglement
preparation that relied on the unitary or dissipative dynamics
and where the entanglement infidelities were lower-bounded by the
system
cooperativities~\cite{sorensen2003measurement,kastoryano2011dissipative,shen2011steady,lin2013dissipative,su2014scheme,reiter2016scalable},
our approach can, in principle, make the entanglement infidelity
arbitrarily small by increasing the squeezing parameter of the
cavity mode for a modest value of the cooperativity.

\begin{figure}[tbph]
\centering
\includegraphics[width=8.0cm]{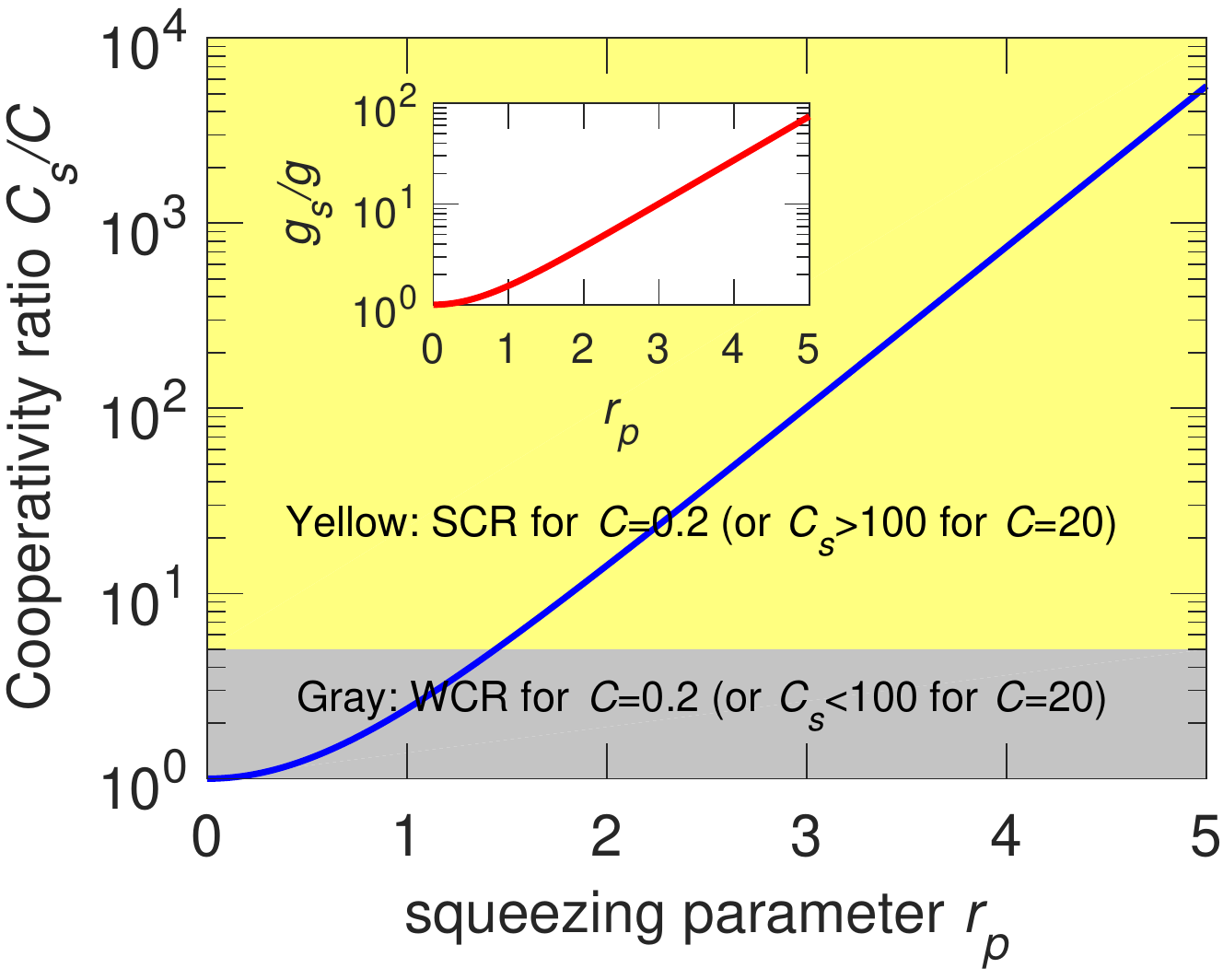}
\caption{Cooperativity enhancement $C_{s}/C$ versus
the squeezing parameter $r_{p}$. For $C=0.2$, the gray and yellow
shaded areas represent the WCR ($C_{s}<1$) and the SCR
($C_{s}>1$), respectively. For $C=20$, the two shaded areas
represent the regions, respectively, with $C_{s}<100$ and
$C_{s}>100$. The inset shows the exponentially-enhanced effective
coupling, $g_{s}$, between atom and cavity.}\label{figenhcoop}
\end{figure}

\emph{Enhanced light-matter interaction and cooperativity.}---Specifically, in a proper observation frame, the Hamiltonian
determining the unitary dynamics of the system reads (hereafter,
we set $\hbar=1$)
\begin{align}\label{eq:full_Hamiltonian}
H\left(t\right)=&\sum_{k}\left(\Delta_{e}|e\rangle_{k}\langle e|+\Delta_{f}|f\rangle_{k}\langle f|\right)+H_{\text{NL}}+H_{\text{AC}}\nonumber\\
&+\frac{1}{2}\Omega_{\text{MW}}\sum_{k}\left(|f\rangle_{k}\langle
g|+\text{H.c.}\right)+V\left(t\right).
\end{align}
Here, $k=1,2$ labels the atoms,
$H_{\text{NL}}=\Delta_{c}a^{\dag}a+\frac{1}{2}\Omega_{p}\left(e^{i\theta_{p}}a^{2}+\text{H.c.}\right)$
is the nonlinear Hamiltonian for degenerate parametric
amplification, $H_{\text{AC}}=g\sum_{k}\left(a|e\rangle_{k}\langle
f|+\text{H.c.}\right)$ is the atom-cavity coupling Hamiltonian,
and $V\left(t\right)=\frac{1}{2}\Omega e^{i\beta
t}\sum_{k}\left[\left(-1\right)^{k-1}|g\rangle_{k}\langle
e|+\text{H.c.}\right]$ describes the interaction of a classical
laser drive with the atoms. The detunings are
$\Delta_{e}=\omega_{e}-\omega_{g}-\omega_{\text{MW}}-\omega_{p}/2$,
$\Delta_{f}=\omega_{f}-\omega_{g}-\omega_{\text{MW}}$, and
$\beta=\omega_{L}-\omega_{\text{MW}}-\omega_{p}/2$, where
$\omega_{L}$ is the laser frequency of the atom drive and
$\omega_{z}$ is the frequency associated with level $|z\rangle$
($z=g,f,e$). Upon introducing the Bogoliubov squeezing
transformation
$a_{s}=\cosh\left(r_{p}\right)a+\exp\left(-i\theta_{p}\right)\sinh\left(r_{p}\right)a^{\dag}$~\cite{scully1997book},
$H_{\text{NL}}$ is diagonalized to
$H_{\text{NL}}=\omega_{s}a_{s}^{\dag}a_{s}$, where
$\omega_{s}=\Delta_{c}\sqrt{1-\alpha^2}$ is the squeezed-cavity
frequency. The atom-cavity coupling Hamiltonian likewise becomes
$H_{\text{AC}}=\sum_{k}\left[\left(g_{s}a_{s}-g^{\prime}_{s}a_{s}^{\dag}\right)|e\rangle_{k}\langle
f|+\text{H.c.}\right]$, with $g_{s}=g\cosh\left(r_{p}\right)$ and
$g^{\prime}_{s}=\exp\left(-i\theta_{p}\right)g\sinh\left(r_{p}\right)$.
The excitation-number-nonconserving processes originating from the
counter-rotating terms of the form
$a_{s}^{\dag}\sum_{k}|e\rangle_{k}\langle f|$, and
$a_{s}\sum_{k}|f\rangle_{k}\langle e|$ can be neglected under the
assumption that
$|g^{\prime}_{s}|/\left(\omega_{s}+\Delta_{e}-\Delta_{f}\right)\ll1$,
corresponding to the rotating-wave approximation, such that
$H_{\text{AC}}$ is transformed to the Jaynes-Cummings Hamiltonian
\begin{align}
H_{\text{ASC}}=g_{s}\sum_{k}\left(a_{s}|e\rangle_{k}\langle
f|+\text{H.c.}\right),
\end{align}
given in terms of the coupling strength $g_{s}$ between the atoms
and the squeezed-cavity mode. Therefore for $r_{p}\geq1$, we
predict an \emph{exponentially-enhanced atom-cavity coupling},
\begin{equation}\label{enhancement1}
  \frac{g_{s}}{g}\sim \frac{1}{2}\exp\left(r_{p}\right),
\end{equation}
as plotted in the inset of Fig.~\ref{figenhcoop}. This is because
there are $\sim \exp\left(2r_{p}\right)$ photons converted into a
single-photon state, $|1\rangle_{s}$, of the squeezed-cavity mode.
Such an exponential enhancement of this light-matter interaction
is one of our most important results.

This squeezing also introduces additional noise into the cavity,
as mentioned in the description above. To circumvent such
undesired noises, a squeezed-vacuum field, with a squeezing
parameter $r_{e}$ and a reference phase $\theta_{e}$, is used to
drive the cavity [see Fig.~\ref{figschematic}(a)]. We consider the
case where such a field has a much larger linewidth than the
cavity mode. Indeed, a squeezing bandwidth of up to $\sim$ GHz has
been experimentally demonstrated via optical parametric
amplification~\cite{ast2013high, serikawa2016creation,
ockeloen2017noiseless}. Because the linewidth is $\sim$ MHz for
typical optical cavities, we can think of this cavity drive as a
squeezed reservoir. Hence, by ensuring $r_{e}=r_{p}$ and
$\theta_{e}+\theta_{p}=\pm n\pi$ ($n=1,3,5,\cdots$), this
additional noise can be eliminated completely (see the
Supplemental Material~\cite{supplement} for details). As a
consequence, the squeezed-cavity mode is equivalently coupled to a
thermal vacuum reservoir, so that we can use the standard Lindblad
operator to describe the cavity decay, yielding
$L_{\text{as}}=\sqrt{\kappa}a_{s}$ with $\kappa$ a decay rate.
Similarly, atomic spontaneous emission is also described with the
Lindblad operators
$L_{g1}=\sqrt{\gamma_{g}}|g\rangle_{1}\langle e|$,
$L_{f1}=\sqrt{\gamma_{f}}|f\rangle_{1}\langle e|$,
$L_{g2}=\sqrt{\gamma_{g}}|g\rangle_{2}\langle e|$, and
$L_{f2}=\sqrt{\gamma_{f}}|f\rangle_{2}\langle e|$. Here, we have
assumed that in each atom, $|e\rangle$ decays to $|g\rangle$ and
$|f\rangle$, respectively, with rates $\gamma_{g}$ and
$\gamma_{f}$. The dynamics of the atom-cavity system is, thus,
governed by the standard master equation in the Lindblad form
$\dot{\rho}\left(t\right)=i\left[\rho\left(t\right),H_{s}\left(t\right)\right]-\frac{1}{2}\sum_{n}\mathcal{L}\left(L_{n}\right)\rho\left(t\right)$,
where $\rho\left(t\right)$ is the density operator of the system,
$H_{s}\left(t\right)$ is given by $H\left(t\right)$ but with $a$
($a^\dagger$) replaced by $a_{s}$ ($a_s^\dagger$), and with
$H_{\text{AC}}$ replaced by $H_{\text{ASC}}$. Moreover,
$\mathcal{L}\left(o\right)\rho= o^{\dag}o\rho-2o\rho o^{\dag}+\rho
o^{\dag}o$ and the sum runs over all dissipative processes
mentioned above. We find that the above master equation gives an
effective cooperativity
$C_{s}=g_{s}^{2}/\left(\kappa\gamma\right)$. Consequently,
increasing $r_{p}$ enables an exponential enhancement in the
atom-cavity coupling, given in Eq.~(\ref{enhancement1}), and
thus, the cooperativity enhancement,
\begin{align}\label{enhancement}
\frac{C_{s}}{C}\sim\frac{1}{4}\exp\left(2r_{p}\right),
\end{align}
as shown in Fig.~\ref{figenhcoop}. Note that our approach can
exponentially strengthen the coherent coupling between atom and
cavity, but \emph{does not introduce any additional noise} into
the system. It is seen in Fig.~\ref{figenhcoop} that the
atom-cavity system can be driven from the weak-coupling regime
(WCR) to the SCR, e.g., with $C=0.2$ and $r_{p}\geq1.5$. Moreover,
an effective cooperativity of $C_{s}>10^2$ can also be achieved
with modest $C$ and $r_{p}$, e.g., $C=20$ and $r_{p}\geq1.5$. As
one of many possible applications in quantum information
technologies, this enhancement in the cooperativity (or coherent
atom-field coupling) can be employed to improve the fidelity of
dissipative entanglement preparation.

\begin{figure}
\centering
\begin{minipage}[c]{4.25cm}
    \centering
    \includegraphics[width=4.25cm,angle=0]{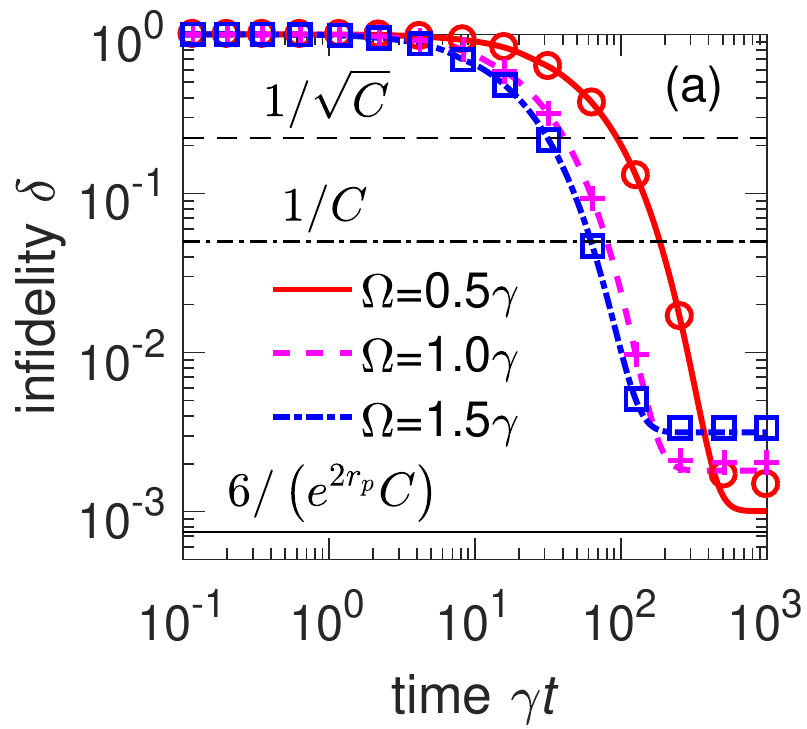}
\end{minipage}
\begin{minipage}[c]{4.25cm}
    \centering
    \includegraphics[width=4.25cm,angle=0]{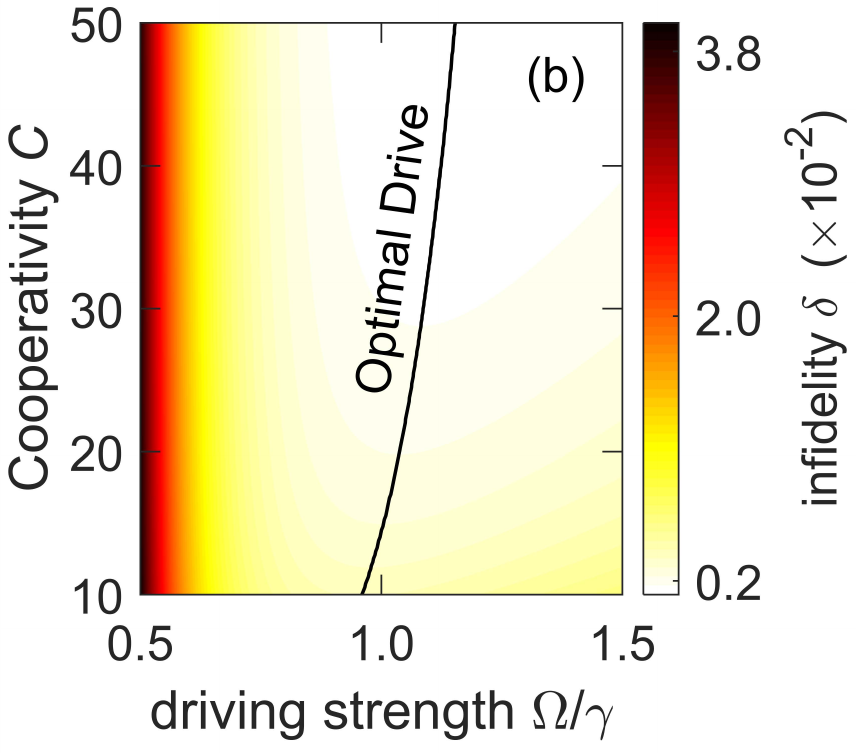}
\end{minipage}
\caption{(a) Evolution of the entanglement
infidelity $\delta$ for different driving strengths
$\Omega=0.5\gamma$, $1.0\gamma$, and $1.5\gamma$, with the
cooperativity $C=20$. We assumed $\Delta_{f}=\Omega/2^{7/4}$ and
$\Delta_{f}=\Omega/2^{7/4}+|g_{s}^{\prime}|^2/\left(2\Delta_{e}\right)$,
$\Delta_{e}=200g_{s}^{\prime}$ when using the effective (thick
curves) and full (symbols) master equations, respectively. This
yields an excellent agreement especially for time $t\in[0,
500/\gamma]$. The steady-state error decreases as $\Omega$ and
becomes closer to $6/\left(e^{2r_{p}}C\right)$ (thin solid line),
far below both $1/\sqrt{C}$ (thin dashed line) and $1/C$ (thin
dotted-dashed line). (b) Entanglement infidelity at $t=200/\gamma$
as a function of $C$ and $\Omega$. Here, due to excellent
agreement between our predictions based on the full and effective
master equations in panel (a), only the latter equation was used
in panel (b). The solid line represents the optimal drive
resulting in the smallest error for a given cooperativity. In both
plots, we have assumed that $\gamma_{g}=\gamma/2$,
$\kappa=2\gamma/3$, $\Omega_{\text{MW}}=\sqrt{2}\Delta_{f}$,
$r_{p}=3$, $\theta_{p}=\pi$, while the initial state of the atoms
is $\left(\mathcal{I}-|\psi_{-}\rangle\langle\psi_{-}|\right)/3$
and the cavity is initially in the vacuum.}\label{error}
\end{figure}

\emph{Maximizing steady-state entanglement.}---Let us consider a
weak drive $\Omega$, so that the dominant dynamics of the system
is restricted to a subspace having, at most, one excitation and we
can treat $V\left(t\right)$ as a perturbation to the
system~\cite{reiter2012effective}. After adiabatically eliminating
the excited states, the effective Hamiltonian is given by
$H_{\text{eff}}=\Delta_{f}\left(\mathcal{I}/2-|\phi_{+}\rangle\langle\phi_{-}|+\text{H.c.}\right)+\Omega_{\text{MW}}\left(|\psi_{+}\rangle\langle
\phi_{+}|+\text{H.c.}\right)$, where $\mathcal{I}$ is an identity
operator acting on the ground manifold of the atoms. This implies
that the microwave field can drive the population from
$|\phi_{+}\rangle$ (or $|\phi_{-}\rangle$) to $|\psi_{+}\rangle$.
Upon choosing $\Delta_{e}=\beta=\omega_{s}+\Delta_{f}$, the
population in $|\psi_{+}\rangle$ is transferred to
$|\psi_{-}\rangle$ via the resonant drive and then the atomic
spontaneous emission, which is mediated by the dark state
$|D\rangle$. At the same time, the transition from
$|\psi_{-}\rangle$ to the excited state of
$|\varphi_{e}\rangle=\left(|fe\rangle+|ef\rangle\right)|0\rangle_{s}/\sqrt{2}$
is off-resonant, and it is negligible when $\Omega\ll g_{s}$. In this
case, the rates of the effective decays into and out of the
desired state $|\psi_{-}\rangle=|S\rangle|0\rangle_{s}$ are
expressed, respectively, as
$\Gamma_{\text{in}}=\left(\Omega/2\right)^{2}\left[4\gamma_{g}/\gamma^2+4/\left(\gamma
C_{s}\right)+\gamma_{f}/(2\gamma^2C_{s}^{2})\right]$ and
$\Gamma_{\text{out}}=\left(\Omega/2\right)^{2}\left[1/\left(\gamma
C_{s}\right)+\left(\gamma+\gamma_{f}\right)/\left(16\gamma^{2}C_{s}^{2}\right)\right]$
(see the Supplemental Material~\cite{supplement} for a detailed
derivation). Here, $\gamma=\gamma_{g}+\gamma_{f}$ is the total
atomic decay rate. In the steady state, the entanglement infidelity
can be expressed as $\delta\sim
1/\left[1+\Gamma_{\text{in}}/\left(3\Gamma_{\text{out}}\right)\right]$,
which is reduced to
$\delta\sim3\gamma/\left(4\gamma_{g}C_{s}\right)$ for $C_{s}\gg1$.
Further, as long as $r_{p}\geq1$, we directly obtain
\begin{align}
\delta\sim\frac{3\gamma}{\gamma_{g}\exp\left(2r_{p}\right)C}.
\end{align}
This explicitly shows an exponential improvement over the
infidelity in the case of previous entanglement preparation
protocols relying on engineered dissipation. The
parametrically-enhanced cooperativity enables the entanglement
infidelity to be very close to zero even for a modest value of
$C$, rather than lower-bounded by $1/\sqrt{C}$ and $1/C$ [see
Fig.~\ref{error}(a)]. For the cooperativity values, which are
easily accessible in current experiments, an entanglement
infidelity of up to $\delta\sim10^{-3}$ can be generated at a time
$t=200/\gamma$, as shown in Fig.~\ref{error}(b). Note that, by
increasing the driving laser strength $\Omega$, the population
transfer into the desired state is faster and, then, the
infidelity is smaller for a given preparation time. However, at the
same time, a nonadiabatic error increases with $\Omega$, causing
an increase in the infidelity. Thus, these are two competing
processes. In addition, a larger $C$ can more strongly reduce this
nonadiabatic error and, therefore, lead to a smaller optimal
driving strength [see Fig. \ref{error}(b)]. In a realistic setup
based on ultracold $^{87}$Rb atoms coupled to a Fabry-Perot
resonator as discussed below~\cite{hamsen2017two}, an atomic
linewidth of $\gamma/2\pi=3$ MHz and the cooperativity of $C=42$
could result in $\delta\sim1.2\times10^{-3}$, together with
$t\sim11~\mu$s, which allows us to neglect atomic decoherence.

We now consider the counter-rotating terms. In the limit
$|g_{s}^{\prime}|/\Delta_e\ll1$, we find that such terms cause an
energy shift of $|g_{s}^{\prime}|^2/\left(2\Delta_{e}\right)$ to
be imposed on the ground states and a coherent coupling, of
strength $|g_{s}^{\prime}|^2/\left(2\Delta_{e}\right)$, between
the states $|\phi_{+}\rangle$ and
$|\phi_{-}\rangle$~\cite{gamel2010time}. To remove these
detrimental effects, the detunings need to be modified as
$\Delta_{e}=\beta-|g_{s}^{\prime}|^{2}/\left(2\Delta_{e}\right)=\omega_{s}+\Delta_{f}-|g_{s}^{\prime}|^{2}/\Delta_{e}$
and
$\Delta_{f}=\Omega_{\text{MW}}/\sqrt{2}+|g_{s}^{\prime}|^{2}/\left(2\Delta_{e}\right)$,
according to the analysis given in the Supplemental
Material~\cite{supplement}. In this situation, the full system can
be mapped to a simplified system that excludes the
counter-rotating terms and has been discussed above. We
numerically integrate the full master equation with the modified
detunings~\cite{johansson2012qutip,johansson2013qutip2}, and find
that, as in Fig.~\ref{error}(a), the exact entanglement infidelity
is in excellent agreement with the prediction of the effective
dynamics during a very long time interval (e.g., $0\leq
t\leq500/\gamma$).

\emph{Possible implementations.}---We consider a possible
experimental implementation utilizing ultracold $^{87}$Rb atoms
trapped in a high-finesse Fabry-Perot
resonator~\cite{hamsen2017two}. Here, the $^{87}$Rb atoms are used
for the $\Lambda$-configuration atoms and the Fabry-Perot
resonator works as the single-mode cavity. When focusing on
electric-dipole transitions of the $D_1$ line at a wavelength of
$795$ nm, we choose $|g\rangle\equiv|F=1, m_{F}=-1\rangle$,
$|f\rangle\equiv|F=2, m_{F}=-2\rangle$, and
$|e\rangle\equiv|F^{\prime}=2, m^{\prime}_{F}=-2\rangle$, where
$F^{\left(\prime\right)}$ and $m_{F}^{\left(\prime\right)}$ are
quantum numbers characterizing the Zeeman states in the manifolds
$5S_{1/2}$ ($5P_{1/2}$). In this situation, a circularly
$\sigma^{-}$-polarized control laser and a $\pi$-polarized-cavity
mode are needed to couple the transitions $|F=1,
m_{F}=-1\rangle\rightarrow$ and $|F=2, m_{F}=-2\rangle\rightarrow
|F^{\prime}=2, m^{\prime}_{F}=-2\rangle$, respectively. For the
two ground states, although their electric-dipole transition is
forbidden due to their same parity, a microwave field could
directly couple these states through the magnetic-dipole
interaction. Such a coupling has experimentally reached values of
hundreds of
kHz~\cite{treutlein2006quantum,sarkany2014controlling}. Moreover,
the cavity mode can be squeezed typically using, e.g., a
periodically-poled KTiOPO$_4$ (PPKTP) crystal~\cite{vahlbruch2008observation,vahlbruch2016detection,schnabel2017squeezed}. In order to
generate a squeezed-vacuum reservoir, we can also use a PPKTP
crystal with a high-bandwidth pump, so the squeezing bandwidth
of up to $\sim$ GHz~\cite{ast2013high,serikawa2016creation} is
possible.

Solid-state implementations can be considered in the context of
nitrogen-vacancy (NV) centers in diamond with a
whispering-gallery-mode (WGM)
microresonator~\cite{park2006cavity}. In this setup, the
electronic spin states of the NV centers are used to form the
$\Lambda$-configuration structures, such that
$|g\rangle\equiv|^{3}A_{2}, m_{s}=-1\rangle$,
$|f\rangle\equiv|^{3}A_{2}, m_{s}=+1\rangle$, and
$|e\rangle\equiv\left(|E_{-},m_{s}=+1\rangle+|E_{+},m_{s}=-1\rangle\right)/\sqrt{2}$.
The NV spins have extremely long coherence times at room
temperature, while the WGM microresonators made out of nonlinear
crystals exhibit strong optical
nonlinearities~\cite{furst2011quantum,sedlmeir2017polarization}.
These are the key requirements for the entanglement preparation
with a weak atom drive and a squeezed-cavity mode.

As an alternative example of solid-state system, the proposed
method of maximizing steady-state entanglement can also be
realized in superconducting quantum
circuits~\cite{you2005superconducting,valenzuela2006microwave,you2008simultaneous},
where two flux or transmon qubits and a coplanar waveguide
resonator are
used~\cite{you2011atomic,yamamoto2014superconducting}. A
superconducting quantum interference device (SQUID) can be
inserted into the resonator, which is able to create the squeezed
vacuum in the
resonator~\cite{moon2005theory,zagoskin2008controlled,
zhong2013squeezing,murch2013reduction,kono2017nonclassical,bienfait2017magnetic}.
All required parts of such devices have been implemented in
superconducting experiments~\cite{gu2017microwave}.

\emph{Conclusions.}---We have shown that parametric squeezing
enables an exponential enhancement of both: coherent coupling
between an atom and a cavity, as well as the corresponding
cooperativity. As a simple application, the steady-state
entanglement preparation, which results in an exponentially better
fidelity than previous dissipation-based protocols, has also been
demonstrated here. In principle, our method can be extended to
other local quantum operations, e.g., many-body entanglement
preparation~\cite{amico2008entanglement,reiter2016scalable} and
quantum gate implementations~\cite{borregaard2015heralded,
duan2004scalable,majer2007coupling, reiserer2014quantum,
hacker2016photon}. We suggest to use squeezed light for only
performing local intracavity quantum operations and to turn it
off for converting stationary qubits into flying qubits. Moreover,
due to a controllable squeezed-cavity frequency, the present
method should enable reaching the ultra-SCR in optical cavities.
Thus, one may observe many interesting phenomena in cavity-QED,
similar to those observed in circuit
QED~\cite{gu2017microwave,garziano2016one,kockum2017deterministic,stassi2017quantum}.
Indeed, in particular for optical cavities, enhancing the light-matter
interaction and cooperativities is of both fundamental and
practical importance, so we expect that this technique could find
diverse applications in quantum
technologies~\cite{Buluta2011,reiserer2015cavity}.

\begin{acknowledgments}

W.Q. and J.Q.Y. were supported in part by the National Key
Research and Development Program of China (Grant No.
2016YFA0301200), the China Postdoctoral Science Foundation (Grant
No. 2017M610752), the MOST 973 Program of China (Grant No.
2014CB921401), the NSFC (Grant No. 11774022 ), and the NSAF (Grant
No. U1530401). A.M. and F.N. acknowledge the support of a grant
from the John Templeton Foundation. F.N. was partially supported
by the MURI Center for Dynamic Magneto-Optics via the AFOSR Award
No. FA9550-14-1-0040, the Japan Society for the Promotion of
Science (KAKENHI), the IMPACT program of JST, CREST Grant No.
JPMJCR1676, RIKEN-AIST Challenge Research Fund, and JSPS-RFBR
Grant No. 17-52-50023.

\end{acknowledgments}


%


\clearpage \widetext
\begin{center}
\section{\Large Supplemental Material}
\end{center}
\setcounter{equation}{0} \setcounter{figure}{0}
\setcounter{table}{0} \setcounter{page}{1} \makeatletter
\renewcommand{\theequation}{S\arabic{equation}}
\renewcommand{\thefigure}{S\arabic{figure}}
\renewcommand{\bibnumfmt}[1]{[S#1]}
\renewcommand{\citenumfont}[1]{S#1}

\begin{quote}
In this Supplemental Material to the article on
``Exponentially-Enhanced Light-Matter Interaction,
Cooperativities, and Steady-State Entanglement Using Parametric
Amplification'', we first present more details of the elimination
of squeezing-induced noises to show an exponential enhancement of
the light-matter interaction, as well as of the cooperativity.
Then, we derive an effective master equation including an
effective Hamiltonian and effective Lindblad operators, and also
give a detailed description of our entanglement preparation
method. Finally, we discuss, in detail, the effects of
counter-rotating terms and show how to remove them.
\end{quote}
\section{Elimination of squeezing-induced fluctuation noise}

To demonstrate more explicitly the elimination of the
squeezing-induced noise, we now derive the Lindblad master
equation for our atom-cavity system. In addition to an exponential
enhancement of the atom-cavity coupling, the squeezing can
introduce undesired noise, including thermal noise and two-photon
correlations, into the cavity mode. In order to avoid such noises,
our approach employs an auxiliary, high-bandwidth squeezed-vacuum
field, which can be experimentally generated, e.g., via optical
parametric
amplification~\cite{Xast2013high,Xserikawa2016creation}. Owing to
the bandwidth of the squeezed-vacuum field of up to $\sim$ GHz,
the auxiliary field can be thought of as a squeezed-vacuum
reservoir for a typical cavity mode with its bandwidth of order of
MHz. When being coupled to the cavity mode, the auxiliary field
can suppress or even completely eliminate these undesired types of
noise of the squeezed-cavity mode.

The Hamiltonian determining the unitary dynamics of our
atom-cavity system, as shown in Fig.~1, is given by Eq.~(1) and,
for convenience, is recalled here
\begin{align}
\label{seq:full_Hamiltonian}
H\left(t\right)=\;&\sum_{k}\left[\Delta_{e}|e\rangle_{k}\langle e|+\Delta_{f}|f\rangle_{k}\langle f|\right]+H_{\text{AC}}+H_{\text{NL}}\nonumber\\
&+\frac{1}{2}\Omega_{\text{MW}}\sum_{k}\left(|f\rangle_{k}\langle g|+\text{H.c.}\right)+V\left(t\right),\\
\label{seq:DPA}
H_{\text{NL}}=\;&\Delta_{c}a^{\dag}a+\frac{1}{2}\Omega_{p}\left[\exp\left(i\theta_{p}\right)a^{2}+\text{H.c.}\right],\\
\label{seq:atom_cavity_coupling}
H_{\text{AC}}=\;&g\sum_{k}\left(a|e\rangle_{k}\langle f|+\text{H.c.}\right),\\
\label{seq:laserdrive}
V\left(t\right)=\;&\frac{1}{2}\Omega\exp\left(i\beta
t\right)\sum_{k}\left[\left(-1\right)^{k-1}|g\rangle_{k}\langle
e|+\text{H.c.}\right].
\end{align}
Here $k=1,2$ labels the atoms, $g$ is the atom-cavity coupling,
the annihilation operator $a$ corresponds to the cavity mode,
$\Omega$ ($\Omega_{\text{MW}}$) is the Rabi frequency of the laser
(microwave) drive applied to the atoms, and $\Omega_{p}$
($\theta_{p}$) is the amplitude (phase) of the strong pump applied
to the nonlinear medium. We have defined the following detunings:
\begin{align}
\Delta_{c}=\;&\omega_{c}-\omega_{p}/2,\\
\Delta_{e}=\;&\omega_{e}-\omega_{g}-\omega_{\text{MW}}-\omega_{p}/2,\\
\Delta_{f}=\;&\omega_{f}-\omega_{g}-\omega_{\text{MW}},\\
\beta=\;&\omega_{L}-\omega_{\text{MW}}-\omega_{p}/2,
\end{align}
where $\omega_{c}$ is the cavity frequency, $\omega_{L}$
($\omega_{\text{MW}}$) is the frequency of the laser (microwave)
drive applied to the atoms, $\omega_{p}$ is the frequency of the
strong pump applied to the nonlinear medium, and $\omega_{z}$ is
the frequency associated with level $|z\rangle$ ($z=g,f,e$). When
the cavity mode is coupled to the squeezed-vacuum reservoir with a
squeezing parameter $r_{e}$ and a reference phase $\theta_{e}$,
the dynamics of the atom-cavity system is described by the
following master equation~\cite{Xscully1997book}:
\begin{align}\label{Seq:full_masterequation}
\dot{\rho}\left(t\right)=&i\left[\rho\left(t\right),H\left(t\right)\right]
-\frac{1}{2}\Bigg\{\sum_{x'}\mathcal{L}\left(L_{x'}\right)\rho\left(t\right)
+\left(N+1\right)\mathcal{L}\left(L_{a}\right)\rho\left(t\right)\nonumber\\
&+N\mathcal{L}\left(L_{a}^{\dag}\right)\rho\left(t\right)
-M\mathcal{L}^{\prime}\left(L_{a}\right)\rho\left(t\right)-M^{*}\mathcal{L}^{\prime}\left(L_{a}^{\dag}\right)\rho\left(t\right)\Bigg\},
\end{align}
where $\rho\left(t\right)$ is the density operator of the system,
a Lindblad operator $L_{a}=\sqrt{\kappa}a$ describes the cavity
decay with a rate $\kappa$, and
\begin{equation}
  N=\sinh^{2}\left(r_{e}\right)\quad \text{and}\quad M=\cosh\left(r_{e}\right)\sinh\left(r_{e}\right)e^{-i\theta_{e}}
 \label{N11}
\end{equation}
describe thermal noise and two-photon correlations caused by the
squeezed-vacuum reservoir, respectively. Moreover,
\begin{eqnarray}
  \mathcal{L}\left(o\right)\rho\left(t\right)&=&o^{\dag}o\rho\left(t\right)-2o\rho\left(t\right) o^{\dag}+\rho\left(t\right)
  o^{\dag}o, \\
  \mathcal{L}'\left(o\right)\rho\left(t\right)&=&oo\rho\left(t\right)-2o\rho\left(t\right)
o+\rho\left(t\right) oo \label{N12}
\end{eqnarray}
and the sum runs over all atomic spontaneous emissions, including
the Lindblad operators
\begin{equation}
  L_{g1}=\sqrt{\gamma_{g}}|g\rangle_{1}\langle e|,\quad
  L_{f1}=\sqrt{\gamma_{f}}|f\rangle_{1}\langle e|,\quad
  L_{g2}=\sqrt{\gamma_{g}}|g\rangle_{2}\langle e|,\quad
  L_{f2}=\sqrt{\gamma_{f}}|f\rangle_{2}\langle e|.
 \label{N1}
\end{equation}
Note that, here, we have assumed that the atoms are coupled to a
thermal reservoir and that in each atom, $|e\rangle$ decays to
$|g\rangle$ and $|f\rangle$, respectively, with rates $\gamma_{g}$
and $\gamma_{f}$.

When pumped, the nonlinear medium can squeeze the cavity mode
along the axis rotated at an angle
$\left(\pi-\theta_{p}\right)/2$, with a squeezing parameter
$r_{p}=\left(1/4\right)\ln\left[\left(1+\alpha\right)/\left(1-\alpha\right)\right]$,
where $\alpha=\Omega_{p}/\Delta_{c}$. This results in a
squeezed-cavity mode, as described by the Bogoliubov
transformation
$a_{s}=\cosh\left(r_{p}\right)a+\exp\left(-i\theta_{p}\right)\sinh\left(r_{p}\right)
a^{\dag}$~\cite{Xscully1997book}, such that
\begin{align}\label{seq:freesqueezedmode}
H_{\text{NL}}=\omega_{s}a_{s}^{\dag}a_{s},
\end{align}
where $\omega_{s}=\Delta_{c}\sqrt{1-\alpha^2}$ is the
squeezed-cavity frequency. In terms of the mode $a_{s}$, the
atom-cavity interaction Hamiltonian $H_{\text{AC}}$ in
Eq.~(\ref{seq:atom_cavity_coupling}) is reexpressed as
\begin{align}\label{seq:fullsquzeedmodeandatoms}
H_{\text{AC}}=\sum_{k}\left[\left(g_{s}a_{s}-g^{\prime}_{s}a_{s}^{\dag}\right)|e\rangle_{k}\langle
f|+\text{H.c.}\right],
\end{align}
where $g_{s}=g\cosh\left(r_{p}\right)$ and
$g_{s}^{\prime}=\exp\left(-i\theta_{p}\right)g\sinh\left(r_{p}\right)$.
Under the assumption that
$|g^{\prime}_{s}|/\left(\omega_{s}+\Delta_{e}-\Delta_{f}\right)\ll1$,
we can make the rotating-wave approximation to neglect the
counter-rotating terms, which results in a standard
Jaynes-Cummings Hamiltonian
\begin{align}
H_{\text{ASC}}=g_{s}\sum_{k}\left(a_{s}|e\rangle_{k}\langle
f|+\text{H.c.}\right).
\end{align}
This Hamiltonian describes an interaction between the atoms and
the squeezed-cavity mode, and demonstrate that as long as
$r_{p}\geq1$, there is an exponential enhancement in the
atom-cavity coupling,
\begin{equation}
  \frac{g_{s}}{g}\sim\frac{1}{2}\exp\left(r_{p}\right).
 \label{N14}
\end{equation}
Furthermore, the master equation in
Eq.~(\ref{Seq:full_masterequation}) can accordingly be reexpressed
as
\begin{align}
\dot{\rho}\left(t\right)=\;&i\left[\rho\left(t\right),H_{s}\left(t\right)\right]\nonumber\\
&-\frac{1}{2}\bigg\{\sum_{x'}\mathcal{L}\left(L_{x'}\right)\rho\left(t\right)
+\left(N_{s}+1\right)\mathcal{L}\left(L_{as}\right)\rho\left(t\right)\nonumber\\
&+N_{s}\mathcal{L}\left(L_{as}^{\dag}\right)\rho\left(t\right)
-M_{s}\mathcal{L}^{\prime}\left(L_{as}\right)\rho\left(t\right)-M_{s}^{*}\mathcal{L}^{\prime}\left(L_{as}^{\dag}\right)\rho\left(t\right)\bigg\},\\
\label{seq_reducedH}
H_{s}\left(t\right)=&\sum_{k}\left[\Delta_{e}|e\rangle_{k}\langle e|+\Delta_{f}|f\rangle_{k}\langle f|\right]+\omega_{s}a_{s}^{\dag}a_{s}+H_{\text{ASC}}\nonumber\\
&+\frac{1}{2}\Omega_{\text{MW}}\sum_{k}\left(|f\rangle_{k}\langle
g|+\text{H.c.}\right)+V\left(t\right),
\end{align}
where $N_{s}$ and $M_{s}$ are given, respectively, by
\begin{align}
\label{effective-thermal-noise}
N_{s}=&\cosh^{2}\left(r_{p}\right)\sinh^{2}\left(r_{e}\right)+\sinh^{2}\left(r_{p}\right)\cosh^{2}\left(r_{e}\right)\nonumber\\
&+\frac{1}{2}\sinh\left(2r_{p}\right)\sinh\left(2r_{e}\right)\cos\left(\theta_{e}+\theta_{p}\right),\\
\label{effective-tow-photon-correlation}
M_{s}=&\exp\left(i\theta_{p}\right)\left[\sinh\left(r_{p}\right)\cosh\left(r_{e}\right)+\exp\left[-i\left(\theta_{e}+\theta_{p}\right)\right]
\cosh\left(r_{p}\right)\sinh\left(r_{e}\right)\right]\nonumber\\
&\times\left[\cosh\left(r_{p}\right)\cosh\left(r_{e}\right)+\exp\left[i\left(\theta_{p}+\theta_{e}\right)\right]\sinh\left(r_{e}\right)
\sinh\left(r_{p}\right)\right],
\end{align}
corresponding to an effective thermal noise and two-photon
correlations of the squeezed-cavity mode, and where
$L_{\text{as}}=\sqrt{\kappa}a_{s}$ is a Lindblad operator
corresponding to the decay of the squeezed-cavity mode,
$g_{s}=g\cosh\left(r_{p}\right)$ is the enhanced, controllable
atom-cavity coupling. We have neglected the counter-rotating terms
to obtain the Hamiltonian $H_{s}$. From
Eqs.~(\ref{effective-thermal-noise}) and
(\ref{effective-tow-photon-correlation}), we can, as $r_{e}=0$,
observe the noise caused only by squeezing the cavity mode.
However, when choosing $r_{e}=r_{p}$ and
$\theta_{e}+\theta_{p}=\pm n\pi$ ($n=1,3,5,\cdots$), we have
\begin{align}
N_{s}=M_{s}=0,
\end{align}
so that the master equation is simplified to a Lindblad form,
\begin{align}\label{Seq:simplified_masterequation}
\dot{\rho}\left(t\right)=i\left[\rho\left(t\right),H_{s}\left(t\right)\right]-\frac{1}{2}\sum_{x}\mathcal{L}\left(L_{x}\right)\rho\left(t\right).
\end{align}
Here, the sum runs over all dissipative processes, including
atomic spontaneous emission and squeezed-cavity decay. From
Eq.~(\ref{Seq:simplified_masterequation}), we find that the
squeezed-cavity mode is equivalently coupled to a thermal
reservoir, and the squeezing-induced noises are completely removed
as desired. Therefore, we can define the effective cooperativity
$C_{s}=g_{s}^{2}/\left(\kappa\gamma\right)$, and obtain an
exponential enhancement in the atom-cavity cooperativity
$C=g^{2}/\left(\kappa\gamma\right)$, that is,
\begin{align}
\frac{C_{s}}{C}=\cosh^{2}\left(r_{p}\right)\sim\frac{1}{4}\exp\left(2r_{p}\right).
\end{align}
This can be used to improve the quality of dissipative
entanglement preparation. The resulting entanglement infidelity is
no longer lower-bounded by the cooperativity $C$ of the
atom-cavity system and could be, in principle, made very close to
zero.

Our method is to use a squeezed-vacuum field to suppress the noise
of the squeezed-cavity mode, including thermal noise and
two-photon correlations. This makes the squeezed-cavity mode
equivalently coupled to a thermal-vacuum reservoir. Therefore,
this method only changes the environment of the squeezed-cavity
mode, and cannot cause the cavity mode to violate the Heisenberg
uncertainty principle. To elucidate more explicitly the physics
underlying this effect and to obtain an analytical understanding,
we consider a simple case when the cavity mode is decoupled from
the atoms. In this case, the Hamiltonian only includes the
nonlinear term given in Eq.~(\ref{seq:DPA}). The cavity mode is
then coupled to the squeezed-vacuum reservoir. Following the same
method as before, we can find that the squeezed-cavity mode is
equivalently coupled to a thermal vacuum reservoir. The
corresponding master equation is
\begin{equation}\label{seq:only-cavity-mode-master-equation}
\dot{\rho}\left(t\right)=i\left[\rho\left(t\right),\omega_{s}a_{s}^{\dag}a_{s}\right]
-\frac{\kappa}{2}\left[a^{\dag}_{s}a_{s}\rho\left(t\right)-2a_{s}\rho\left(t\right)
a_{s}^{\dag}+\rho\left(t\right) a_{s}^{\dag}a_{s}\right].
\end{equation}
We now calculate the Heisenberg uncertainty relation of the cavity
mode $a$ evolving according to the master equation given in
Eq.~(\ref{seq:only-cavity-mode-master-equation}). To start, we
define two rotated quadratures at an angle
$\left(\pi-\theta_{p}\right)/2$,
\begin{align}
X_{1}&=\frac{1}{2}\left\{a\exp\left[-i\left(\pi-\theta_{p}\right)/2\right]+a^{\dag}\exp\left[i\left(\pi-\theta_{p}\right)/2\right]\right\},\\
X_{2}&=\frac{1}{2i}\left\{a\exp\left[-i\left(\pi-\theta_{p}\right)/2\right]-a^{\dag}\exp\left[i\left(\pi-\theta_{p}\right)/2\right]\right\}.
\end{align}
In terms of the $a_{s}$ mode, $X_{1}$ and $X_{2}$ can be
reexpressed as
\begin{align}
X_{1}&=x_{1}a_{s}+x_{1}^{\ast}a_{s}^{\dag},\\
X_{2}&=-i\left(x_{2}a_{s}-x_{2}^{\ast}a_{s}^{\dag}\right).
\end{align}
Here,
\begin{align}
x_{1}&=\frac{1}{2}\left\{\exp\left[-i\left(\pi-\theta_{p}\right)/2\right]\cosh\left(r_{p}\right)
-\exp\left[i\left(\pi+\theta_{p}\right)/2\right]\sinh\left(r_{p}\right)\right\},\\
x_{2}&=\frac{1}{2}\left\{\exp\left[-i\left(\pi-\theta_{p}\right)/2\right]\cosh\left(r_{p}\right)
+\exp\left[i\left(\pi+\theta_{p}\right)/2\right]\sinh\left(r_{p}\right)\right\}.
\end{align}
According to the master equation in
Eq.~(\ref{seq:only-cavity-mode-master-equation}), a
straightforward calculation gives
\begin{align}
\left(\Delta X_{1}\right)^2&=\langle X_{1}^{2}\rangle-\langle X_{1}\rangle^{2}\nonumber\\
&=\Big\{y_{1}^{2}\exp\left(-i2\omega_{s}t\right)\left[\langle a_{s}a_{s}\rangle\!\left(0\right)-\langle a_{s}\rangle^{2}\!\left(0\right)\right]\nonumber\\
&+2|y_{1}|^{2}\left[\langle a^{\dag}_{s}a_{s}\rangle\!\left(0\right)-\langle a^{\dag}_{s}\rangle\!\left(0\right)\langle a_{s}\rangle\!\left(0\right)\right]\nonumber\\
&+y_{1}^{\ast2}\exp\left(i2\omega_{s}t\right)\left[\langle a^{\dag}_{s}a^{\dag}_{s}\rangle\!\left(0\right)-\langle a^{\dag}_{s}\rangle^{2}\!\left(0\right)\right]\Big\}\exp\left(-\kappa t\right)+\frac{1}{4}\exp\left(2r_{p}\right),\\
\left(\Delta X_{2}\right)^2&=\langle X_{2}^{2}\rangle-\langle X_{2}\rangle^{2}\nonumber\\
&=\Big\{y_{2}^{2}\exp\left(-i2\omega_{s}t\right)\left[\langle a_{s}\rangle^{2}\!\left(0\right)-\langle a_{s}a_{s}\rangle\!\left(0\right)\right]\nonumber\\
&+2|y_{2}|^{2}\left[\langle a^{\dag}_{s}a_{s}\rangle\!\left(0\right)-\langle a^{\dag}_{s}\rangle\!\left(0\right)\langle a_{s}\rangle\!\left(0\right)\right]\nonumber\\
&+y_{2}^{\ast2}\exp\left(i2\omega_{s}t\right)\left[\langle
a^{\dag}_{s}\rangle^{2}\!\left(0\right)\right]-\langle
a^{\dag}_{s}a^{\dag}_{s}\rangle\!\left(0\right)\Big\}\exp\left(-\kappa
t\right)+\frac{1}{4}\exp\left(-2r_{p}\right),
\end{align}
where $\langle O\rangle\!\left(t\right)$ represents the
expectation value of the operator $O$ at the evolution time $t$.
For simplicity, and without loss of generality, we assume that the
squeezed-cavity mode is initially in a Fock state $|n_{s}\rangle$,
with $n_{s}$ being the squeezed-cavity photon number. In this
case, we have
\begin{align}
\left(\Delta X_{1}\right)^2=&\frac{1}{4}\left[2n_{s}\exp\left(-\kappa t\right)+1\right]\exp\left(2r_{p}\right),\\
\left(\Delta
X_{2}\right)^2=&\frac{1}{4}\left[2n_{s}\exp\left(-\kappa
t\right)+1\right]\exp\left(-2r_{p}\right),
\end{align}
and then
\begin{equation}\label{req:Heisenberg-uncertainty-relation}
\left(\Delta X_{1}\right)\left(\Delta
X_{2}\right)=\frac{1}{4}\left[2n_{s}\exp\left(-\kappa
t\right)+1\right]\geq\frac{1}{4}.
\end{equation}
It is found, from Eq.~(\ref{req:Heisenberg-uncertainty-relation}),
that the Heisenberg uncertainty relation holds, as expected.

We now turn to the discussion of the squeezed vacuum drive. The
squeezing strength $r_{e}$ and squeezing phase $\theta_{e}$ are
experimentally adjustable quantities. In optics, the squeezed
vacuum can be produced by a pumped $\chi^{\left(2\right)}$
nonlinear medium (e.g., a periodically-poled KTiOPO4 (PPKTP)
crystal) placed in an optical
cavity~\cite{Xvahlbruch2016detection, Xserikawa2016creation,
Xast2013high, Xvahlbruch2008observation}. This method is similar
to generating cavity-field squeezing of a atom-cavity system. The
parameters $r_{e}$ and $\theta_{e}$ can be controlled by the
amplitude and phase of the laser, which pumps the crystal. To
confirm the values of the parameters, one can further measure
these by using balanced homodyne
detection~\cite{Xschnabel2017squeezed}. The parameters $r_{p}$ and
$\theta_{p}$ can be controlled analogously in such a way to
fulfill the conditions $r_{e}=r_{p}$ and
$\theta_{e}+\theta_{p}=\pm n\pi$ ($n=1,3,5,\cdots$). We note that
optical squeezing has also been experimentally implemented
utilizing a waveguide cavity~\cite{Xstefszky2017waveguide}.

Superconducting quantum circuits, due to their tunable
nonlinearity and low losses for microwave fields, are other
promising devices for producing squeezed states. The most popular
method to generate microwave squeezing is to use a Josephson
parametric amplifier (JPA)~\cite{Xclark2017sideband,
Xbienfait2017magnetic, Xmurch2013reduction, Xzhong2013squeezing,
Xmallet2011quantum}. The JPA is a superconducting {\it LC}
resonator, which consists of a superconducting quantum
interference device (SQUID). This resonator can be pumped not only
through the resonator, but also by modulating the magnetic flux in
the SQUID. In this case, the parameters $r_{e}$ and $\theta_{e}$
can be controlled by the amplitude and phase of a pump tone used
to modulate the magnetic flux. Recent experiments have shown that
the squeezed vacuum, generated by a JPA, can be used to reduce the
radiative decay of superconducting
qubits~\cite{Xmurch2013reduction} and to modify resonance
fluorescence~\cite{Xtoyli2016resonance}. The squeezing of quantum
noise has also been demonstrated with tunable Josephson
metamaterials~\cite{Xcastellanos2008amplification}.


\section{Perturbative treatment and maximizing steady-state entanglement}
\begin{figure}[tbph]
\centering
\includegraphics[width=17cm]{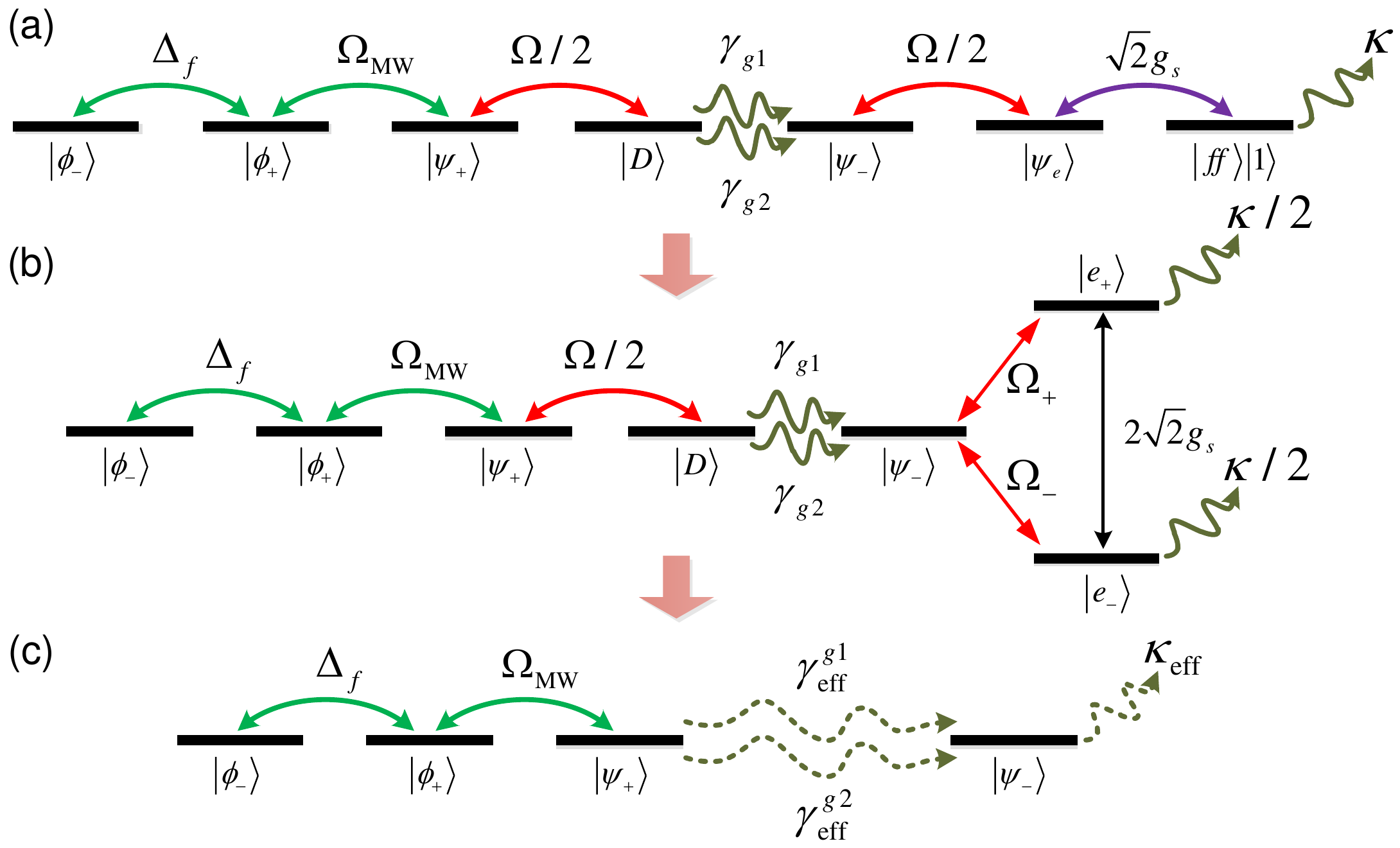}
\caption{(Color online) (a) Hopping-like model for the proposed
steady-state nearly-maximal entanglement preparation. (b)
Exponential suppression in the leakage of the population in
$|\psi_{-}\rangle$. (c) Effective dynamics after adiabatically
eliminating the states $|D\rangle$, $|e_{+}\rangle$, and
$|e_{-}\rangle$.}\label{Sfighopping}
\end{figure}

For the preparation of a steady entangled state, e.g., the singlet
state $|S\rangle=\left(|gf\rangle-|fg\rangle\right)/\sqrt{2}$, the
key element is that the system dynamics cannot only drive the
population into $|\psi_{-}\rangle$, but also prevent the
population from moving out of $|\psi_{-}\rangle$. In our approach,
when we choose $\Delta_{e}=\beta=\omega_{s}+\Delta_{f}$, the
coherent couplings mediated by the laser drive and by the
squeezed-cavity mode are resonant. In addition, the microwave
field also resonantly drives the transition
\begin{equation}
|\phi_{-}\rangle\leftrightarrow|\phi_{+}\rangle\leftrightarrow|\psi_{+}\rangle.
\end{equation}
The proposed entanglement preparation can, therefore, be
understood via a hopping-like model, as illustrated in
Fig.~\ref{Sfighopping}(a). Note that, here, $\Delta_{f}$ is
required to be nonzero, or $|\phi_{-}\rangle$ becomes a dark state
of the microwave drive, whose population is trapped and cannot be
transferred to $|\psi_{+}\rangle$. In the preparation process, the
populations initially in the states $|\phi_{-}\rangle$,
$|\phi_{+}\rangle$, and $|\psi_{+}\rangle$ can be coherently
driven to the dark state $|D\rangle$ through the microwave and
laser drives and, then, decay to the desired state
$|\psi_{-}\rangle$ through two atomic decays, respectively, with
rates $\gamma_{g1}$ and $\gamma_{g2}$. Indeed, such atomic decays
originate, respectively, from the spontaneous emissions,
$|e\rangle\rightarrow|g\rangle$, of the two atoms, so we have
$\gamma_{g1}=\gamma_{g2}=\gamma_{g}/4$. Furthermore, owing to the
laser drive, the state $|\psi_{-}\rangle$ is resonantly excited to
$|\varphi_{e}\rangle$. This state is then resonantly coupled to
$|ff\rangle|1\rangle_{s}$ by the squeezed-cavity mode. The cavity
loss causes the latter state to decay to
$|ff\rangle|0\rangle_{s}$, thus giving rise to population leakage
from $|\psi_{-}\rangle$. However, because of the exponential
enhancement in the atom-cavity coupling [i.e., $g_{s}\sim
g\exp\left(r_{p}\right)/2$ in Eq.~(\ref{N14})], the state
$|\varphi_{e}\rangle$ is split into a doublet of dressed states,
$|e_{\pm}\rangle=\left(|\varphi_{e}\rangle\pm|ff\rangle|1\rangle_{s}\right)/\sqrt{2}$,
exponentially separated by
\begin{equation}
2\sqrt{2}g_{s}\sim\sqrt{2}g\exp\left(r_{p}\right),
\end{equation}
which is much larger than the couplings strength
$\Omega_{\pm}=\Omega/\left(2\sqrt{2}\right)$, as shown in
Fig.~\ref{Sfighopping}(b). Hence, the population leakage from
$|\psi_{-}\rangle$ is exponentially suppressed, and we can make
the effective decay rate, $\Gamma_{\text{out}}$, out of
$|\psi_{-}\rangle$, exponentially smaller than the effective decay
rate, $\Gamma_{\text{in}}$, into $|\psi_{-}\rangle$. To discuss
these decay rates more specifically, we need to give an effective
master equation of the system, when the laser drive $\Omega$ is
assumed to be much smaller than the interactions inside the
excited-state subspace. In this case, the coupling between the
ground- and excited-state subspaces is treated as a perturbation,
so that both cavity mode and excited states of the atoms can be
adiabatically eliminated.

Specifically, we follow the procedure in
Ref.~\cite{Xreiter2012effective}, and begin by considering the
Lindblad master equation in
Eq.~(\ref{Seq:simplified_masterequation}). For convenience, we
rewrite the Hamiltonian $H_{s}\left(t\right)$ as
\begin{align}
H_{s}\left(t\right)=H_{g}+H_{e}+v\!\left(t\right)+v^{\dag}\!\left(t\right),
\end{align}
with
\begin{align}
H_{g}=&\sum_{k=1,2}\left[\Delta_{f}|f\rangle_{k}\langle f|+\frac{\Omega_{\text{MW}}}{2}\left(|f\rangle_{k}\langle g|+\text{H.c.}\right)\right],\\
H_{e}=&\sum_{k=1,2}|e\rangle_{k}\langle
e|+\omega_{s}a_{s}^{\dag}a_{s}+H_{\text{ASC}},
\end{align}
representing the interactions, respectively, inside the ground-
and excited-state subspaces, and
\begin{align}
v\!\left(t\right)=\;&\frac{1}{2}\exp\left(i\beta
t\right)\Omega\sum_{k=1,2}\exp\left[i\left(k-1\right)\pi\right]|g\rangle_{k}\langle
e|
\end{align}
being the deexcitation from the excited-state subspace to the
ground-states subspace. Under the assumption that the laser drive
$\Omega$ is sufficiently weak compared to the coupling $g_{s}$,
the effective Hamiltonian and Lindblad operators read:
\begin{align}
H_{\text{eff}}=\;&-\frac{1}{2}\left[v\!\left(t\right)\left(H_{\text{NH}}-\beta\right)^{-1}v^{\dag}\!\left(t\right)\right]+H_{g},\\
L_{x,
\text{eff}}=\;&L_{x}\left(H_{\text{NH}}-\beta\right)^{-1}v^{\dag}\!\left(t\right),
\end{align}
where
\begin{align}
H_{\text{NH}}=H_{e}-\frac{i}{2}\sum_{x}L_{x}^{\dag}L_{x}
\end{align}
is the no-jump Hamiltonian. The system dynamics is, therefore,
determined by an effective master equation
\begin{align}
\dot{\rho}_{g}\!\left(t\right)=i\left[\rho_{g}\!\left(t\right),H_{\text{eff}}\right]-\frac{1}{2}\sum_{x}\mathcal{L}\left(L_{x,\text{eff}}\right)\rho_{g}\!\left(t\right),
\end{align}
where $\rho_{g}\!\left(t\right)$ is the reduced density operator
associated only with the ground states of the atoms. After a
straightforward calculation restricted in the Hilbert space having
at most one excitation, we have:
\begin{align}
H_{\text{eff}}=\; &\Delta_{f}\left(\mathcal{I}/2-|\phi_{+}\rangle\langle\phi_{-}|+\text{H.c.}\right)+\Omega_{\text{MW}}\left(|\psi_{+}\rangle\langle \phi_{+}|+\text{H.c.}\right),\label{Seq:effHamiltonian}\\
L_{g1,\text{eff}}=\;&r_{g}\left[\left(|\psi_{+}\rangle+|\psi_{-}\rangle\right)\left(\gamma_{\text{eff},0}\langle \psi_{+}|+\gamma_{\text{eff},2}\langle \psi_{-}|\right)+\gamma_{\text{eff},1}\left(|\phi_{+}\rangle+|\phi_{-}\rangle\right)\left(\langle\phi_{+}+\langle\phi_{-}|\right)\right],\label{Seq:efflindbladoperatorg1}\\
L_{g2,\text{eff}}=\;&-r_{g}\left[\left(|\psi_{+}\rangle-|\psi_{-}\rangle\right)\left(\gamma_{\text{eff},0}\langle \psi_{+}|-\gamma_{\text{eff},2}\langle\psi_{-}|\right)+\gamma_{\text{eff},1}\left(|\phi_{+}\rangle+|\phi_{-}\rangle\right)\left(\langle\phi_{+}+\langle\phi_{-}|\right)\right],\label{Seq:efflindbladoperatorg2}\\
L_{f1,\text{eff}}=\;&r_{f}\left[\left(|\phi_{+}\rangle-|\phi_{-}\rangle\right)\left(\gamma_{\text{eff},0}\langle \psi_{+}|+\gamma_{\text{eff},2}\langle\psi_{-}|\right)+\gamma_{\text{eff},1}\left(|\psi_{+}\rangle-|\psi_{-}\rangle\right)\left(\langle\phi_{+}|+\langle\phi_{-}|\right)\right],\label{Seq:efflindbladoperatorf1}\\
L_{f2,\text{eff}}=\;&-r_{f}\left[\left(|\phi_{+}\rangle-|\phi_{-}\rangle\right)\left(\gamma_{\text{eff},0}\langle \psi_{+}|-\gamma_{\text{eff},2}\langle\psi_{-}|\right)+\gamma_{\text{eff},1}\left(|\psi_{+}\rangle+|\psi_{-}\rangle\right)\left(\langle\phi_{+}|+\langle\phi_{-}|\right)\right],\label{Seq:efflindbladoperatorf2}\\
L_{\text{as},\text{eff}}=\;&r_{\text{as}}\left[\kappa_{\text{eff},1}|\psi_{-}\rangle\left(\langle\phi_{+}|+\langle\phi_{-}|\right)-\frac{1}{\sqrt{2}}\kappa_{\text{eff},2}
\left(|\phi_{+}\rangle-|\phi_{-}\rangle\right)\langle
\psi_{-}|\right].\label{Seq:efflindbladoperatoras}
\end{align}
Here,
\begin{align}
\mathcal{I}=\;&|\phi_{+}\rangle\langle\phi_{+}|+|\phi_{-}\rangle\langle\phi_{-}|+|\psi_{+}\rangle\langle\psi_{+}|+|\psi_{-}\rangle\langle\psi_{-}|,\\
|\phi_{\pm}\rangle=\;&\frac{1}{\sqrt{2}}\left(|gg\rangle\pm|ff\rangle\right),\\
|\psi_{\pm}\rangle=\;&\frac{1}{\sqrt{2}}\left(|gf\rangle\pm|fg\rangle\right),
\end{align}
and
\begin{align}
r_{g\left(f\right)}=\;&\exp\left(-i\beta t\right)\frac{\Omega\sqrt{\gamma_{g\left(f\right)}}}{4\gamma},\\
r_{\text{as}}=\;&\exp\left(-i\beta t\right)\frac{\Omega}{2\sqrt{\gamma}},\\
\gamma_{\text{eff},0}=\;&\frac{1}{\widetilde{\Delta}_{e,1}},\\
\gamma_{\text{eff},m}=\;&\frac{\widetilde{\omega}_{s,m}}{\widetilde{\omega}_{s,m}\widetilde{\Delta}_{e,m-1}-mC_{s}},\\
\kappa_{\text{eff},m}=\;&\frac{\sqrt{mC_{s}}}{\widetilde{\omega}_{s,m}\widetilde{\Delta}_{e,m-1}-mC_{s}},
\end{align}
where
\begin{align}
\widetilde{\omega}_{s,m}=&\frac{1}{\kappa}\left(\omega_{s}+m\Delta_{f}-\beta\right)-\frac{i}{2},\\
\widetilde{\Delta}_{e,m-1}=&\frac{1}{\gamma}\left[\Delta_{e}+\left(m-1\right)\Delta_{f}-\beta\right]-\frac{i}{2},
\end{align}
for $m=1,2$, and where $\gamma=\gamma_{g}+\gamma_{f}$ is the total
atomic decay rate.

\begin{figure}[tbph]
\centering
\includegraphics[width=7.0cm]{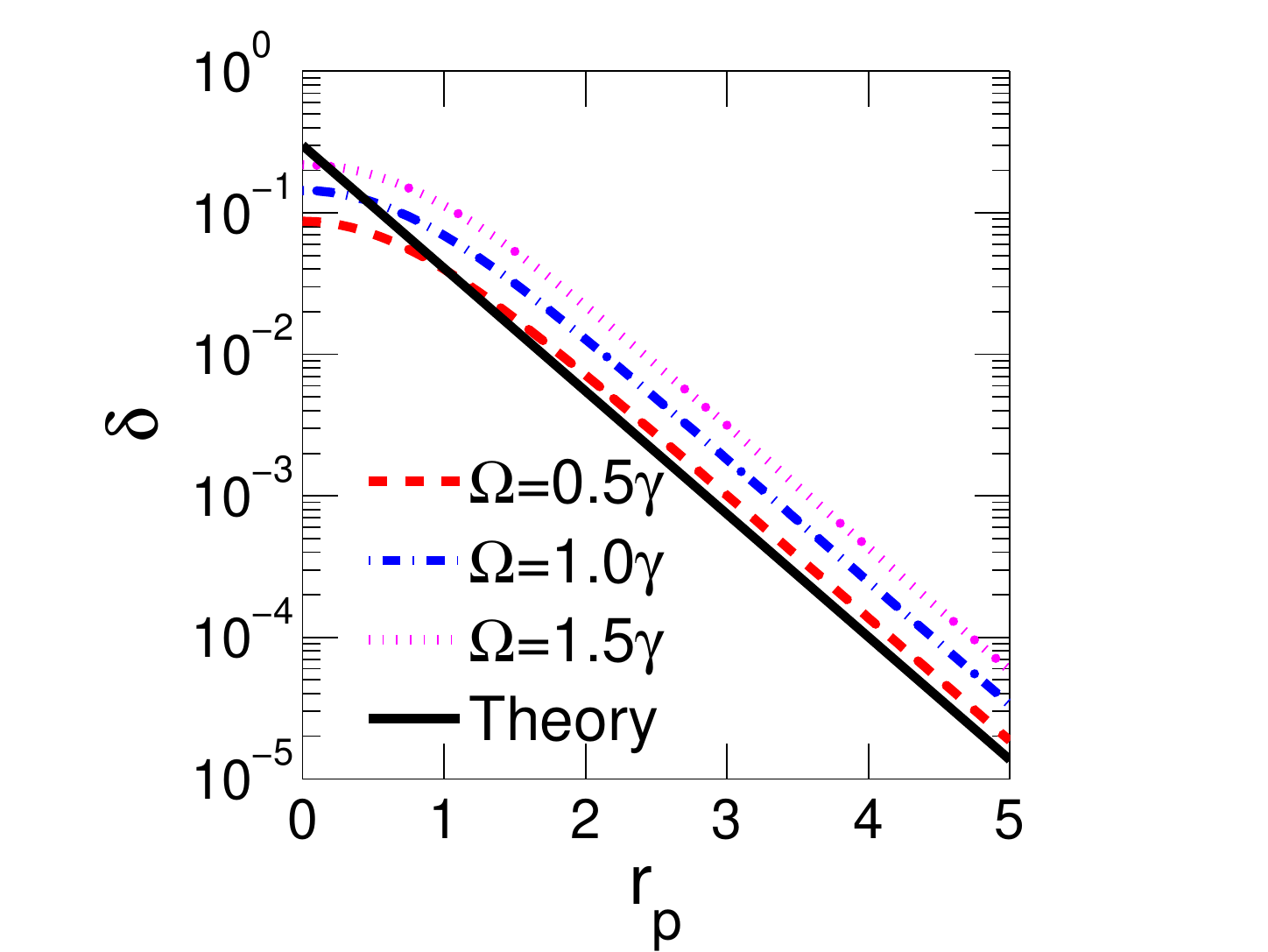}
\caption{(Color online) Steady-state entanglement infidelity
versus the squeezing parameter $r_{p}$. We have plotted the
numerical infidelity for $\Omega=0.5\gamma$ (dashed curve),
$\Omega=1.0\gamma$ (dashed-dotted curve), and $\Omega=1.5\gamma$
(dotted curve) by calculating the effective master equation, and
also plotted the theoretical prediction (solid curve). Here, we
have assumed that $\gamma_{g}=\gamma/2$, $\kappa=2\gamma/3$,
$C=20$, $\Delta_{f}=\Omega/2^{7/4}$,
$\Omega_{\text{MW}}=\sqrt{2}\Delta_{f}$, and that with the vacuum
cavity, the initial state of the atoms is
$\left(\mathcal{I}-|\psi_{-}\rangle\langle\psi_{-}|\right)/3$.}\label{Sfigsteadyerror}
\end{figure}

Having obtained the effective master equation, let us now consider
the decay rates $\Gamma_{\text{in}}$ and $\Gamma_{\text{out}}$.
According to the effective Lindblad operators in
Eqs.~(\ref{Seq:efflindbladoperatorg1})-(\ref{Seq:efflindbladoperatoras}),
the decay rates of moving into and out of the singlet state
$|\psi_{-}\rangle$ are given, respectively, by
\begin{align}
\Gamma_{\text{in}}=\;&\frac{\Omega^{2}}{4\gamma^{2}}\left(\gamma_{g}|\gamma_{\text{eff},0}|^{2}
+2\gamma_{f}|\gamma_{\text{eff},1}|^{2}+4\gamma|\kappa_{\text{eff},1}|^{2}\right),\\
\Gamma_{\text{out}}=\;&\frac{\Omega^{2}}{4\gamma^{2}}\left(\gamma_{g}|\gamma_{\text{eff},2}|^{2}
+2\gamma_{f}|\gamma_{\text{eff},2}|^{2}+2\gamma|\kappa_{\text{eff},2}|^{2}\right).
\end{align}
Let us define the entanglement fidelity as $F=\langle
\psi_{-}|\rho_{g}\left(t\right)|\psi_{-}\rangle$ (that is, the
probability of the atoms being in $|\psi_{-}\rangle$) and, then,
the entanglement infidelity as $\delta=1-F$. In the steady state
($t\rightarrow+\infty$), the entanglement infidelity is found
\begin{align}\label{seq:infidelity0}
\delta\sim\frac{1}{1+\Gamma_{\text{in}}/\left(3\Gamma_{\text{out}}\right)}.
\end{align}
Note that, here, we have assumed that $|\phi_{+}\rangle$,
$|\phi_{-}\rangle$, and $|\psi_{+}\rangle$ have the same
population in a steady state. In order to prepare nearly-maximal
steady-state entanglement, we choose the detunings to be
\begin{align}\label{seq:detuningsforeffmeq}
\Delta_{e}=\beta=\omega_{s}+\Delta_{f},
\end{align}
such that
$\widetilde{\omega}_{s,m}\sim\widetilde{\Delta}_{e,m-1}\sim-i/2$,
yielding
\begin{align}\label{seq:Gammmainout}
\frac{\Gamma_{\text{in}}}{\Gamma_{\text{out}}}\sim\frac{4\gamma_{g}}{\gamma}C_{s}\gg1,
\end{align}
for $C_{s}\gg1$. As shown in Fig.~\ref{Sfighopping}(c), the
underlying dynamics is as follows: after adiabatically eliminating
the excited states $|D\rangle$, $|e_{+}\rangle$, and
$|e_{-}\rangle$, the states $|\psi_{+}\rangle$ and
$|\psi_{-}\rangle$ are directly connected by two effective
spontaneous emission processes with rates
$\gamma^{g1}_{\text{eff}}$ and $\gamma^{g2}_{\text{eff}}$,
\begin{equation}
  \gamma^{g1}_{\text{eff}}=\gamma^{g2}_{\text{eff}}
  =|r_{g}\gamma_{\text{eff},0}|^2\sim\frac{\gamma_{g}}{4\gamma^{2}}\Omega^{2},
 \label{N15}
\end{equation}
and at the same time, the desired state $|\psi_{-}\rangle$ leaks
the population through an effective cavity decay with a rate
$\kappa_{\text{eff}}$,
\begin{equation}
  \kappa_{\text{eff}}=|r_{\text{as}}\kappa_{\text{eff},2}|^2/2\sim\frac{\Omega^{2}}{16\gamma
C_{s}}.
 \label{N16}
\end{equation}
Therefore, together with the effective Hamiltonian
$H_{\text{eff}}$ driving the populations from both
$|\phi_{+}\rangle$ and $|\phi_{-}\rangle$ to $|\psi_{+}\rangle$,
the initial populations in the ground-states subspace of the atoms
can be transferred to $|\psi_{-}\rangle$ and trapped in this
state. By substituting Eq.~(\ref{seq:Gammmainout}) into
Eq.~(\ref{seq:infidelity0}), we can straightforwardly have
\begin{align}
\delta\sim\frac{3\gamma}{4\gamma_{g}C_{s}}.
\end{align}
As long as $r_{p}\geq1$, an exponential enhancement of the
cooperativity, $C_{s}/C\sim \exp\left(2r_{p}\right)/4$, is
obtained, leading to
\begin{align}
\delta\sim\frac{3\gamma}{\gamma_{g}\exp\left(2r_{p}\right)C}.
\end{align}
This equation shows that we can increase the squeezing parameter
$r_{p}$, so as to exponentially decrease the entanglement
infidelity, as seen in Fig.~\ref{Sfigsteadyerror}. Moreover, the
result in this figure also reveals that, by decreasing $\Omega$,
one can suppress non-adiabatic errors and, thus, can cause the
steady-state infidelity to approach a theoretical value, as
expected. Hence, as opposed to prior entanglement preparation
protocols, which relied on controlled unitary dynamics or
engineered dissipation, such an infidelity is no longer lower
bounded by the cooperativity $C$ and, in principle, can be made
very close to zero.

\section{Effects of the counter-rotating terms}
The counter-rotating terms of the form
$a^{\dag}_{s}\sum_{k}|e\rangle_{k}\langle f|$ and
$a_{s}\sum_{k}|f\rangle_{k}\langle e|$, which result from optical
parametric amplification, do not conserve the excitation number,
and can couple the ground- and double-excited states subspaces.
Thus, this would give rise to an additional leakage of the
population in the desired state $|\psi_{-}\rangle$, and decrease
the entanglement fidelity. For example, in the presence of the
counter-rotating terms, the state $|\psi_{-}\rangle$ can be
excited to a double-excitation state
$\left(|ge\rangle-|eg\rangle\right)|1\rangle_{s}/\sqrt{2}$, which,
then, de-excites to the ground state $|gg\rangle|0\rangle$ through
cavity decay and spontaneous emission. In general, we can decrease
the ratio $|g_{s}^{\prime}|/\left(2\Delta_{e}\right)$ to reduce
errors induced by these excitation-number-nonconserving processes.
However, to reduce such errors more efficiently in the limit of
$|g_{s}^{\prime}|/\left(2\Delta_{e}\right)$, we analyze effects of
counter-rotating terms, in detail, in this section, and
demonstrate that by modifying external parameters, we can remove
these terms and the full system can be mapped to a simplified
system described above.

According to Eqs.~(\ref{seq:freesqueezedmode}) and
(\ref{seq:fullsquzeedmodeandatoms}), the full Hamiltonian of the
system in the terms of the squeezed mode $a_{s}$ is
\begin{align}
H\left(t\right)=&\sum_{k}\left[\Delta_{e}|e\rangle_{k}\langle e|+\Delta_{f}|f\rangle_{k}\langle f|\right]+\omega_{s}a_{s}^{\dag}a_{s}\nonumber\\
+&\sum_{k}\left[\left(g_{s}a_{s}-g^{\prime}_{s}a_{s}^{\dag}\right)|e\rangle_{k}\langle f|+\text{H.c.}\right],\nonumber\\
+&\frac{1}{2}\Omega_{\text{MW}}\sum_{k}\left(|f\rangle_{k}\langle g|+\text{H.c.}\right)+V\left(t\right),\\
V\left(t\right)=\;&\frac{1}{2}\Omega\exp\left(i\beta
t\right)\sum_{k}\left[\left(-1\right)^{k-1}|g\rangle_{k}\langle
e|+\text{H.c.}\right].
\end{align}
Indeed, the counter-rotating terms can be treated as the
high-frequency components of the full Hamiltonian. In order to
explicitly show these high-frequency components, we can express
$H\left(t\right)$ into a rotating frame at
\begin{align}
H_{0}=\Delta_{e}\sum_{k}|e\rangle_{k}\langle
e|+\left(\omega_{s}+\Delta_{f}\right)a_{s}^{\dag}a_{s}.
\end{align}
Thus, $H\left(t\right)$ is transformed to
\begin{align}\label{seq:fullHintermediateframe}
\mathcal{H}\left(t\right)=\;&\Delta_{f}\left(\sum_{k}|f\rangle_{k}\langle f|-a_{s}^{\dag}a_{s}\right)\nonumber\\
&+\sum_{k}\left(g_{s}a_{s}|e\rangle_{k}\langle f|-e^{i2\Delta_{e}t}g_{s}^{\prime}a_{s}^{\dag}|e\rangle_{k}\langle f|+\text{H.c.}\right)\nonumber\\
&+\frac{1}{2}\Omega_{\text{MW}}\sum_{k}\left(|f\rangle_{k}\langle g|+\text{H.c.}\right)+\mathcal{V},\\
\mathcal{V}=\;&\frac{1}{2}\Omega\sum_{k}\left[\left(-1\right)^{k-1}|g\rangle_{k}\langle
e|+\text{H.c.}\right].
\end{align}
Here, we have chosen $\Delta_{e}=\beta=\omega_{s}+\Delta_{f}$.
Because $\Delta_{f}$ is required to be much smaller than
$\Delta_{e}$, $\mathcal{H}\left(t\right)$ can be divided into two
parts, $\mathcal{H}\left(t\right)=H_{\text{low}}+H_{\text{high}}$,
where
\begin{align}
H_{\text{low}}=\;&\Delta_{f}\left(\sum_{k}|f\rangle_{k}\langle f|-a_{s}^{\dag}a_{s}\right)+g_{s}\sum_{k}\left(a_{s}|e\rangle_{k}\langle f|+\text{H.c.}\right)\nonumber\\
&+\frac{1}{2}\Omega_{\text{MW}}\sum_{k}\left(|f\rangle_{k}\langle g|+\text{H.c.}\right)+\mathcal{V},\\
H_{\text{high}}=&\sum_{k}\left(-e^{i2\Delta_{e}t}g_{s}^{\prime}a_{s}^{\dag}|e\rangle_{k}\langle
f|+\text{H.c.}\right),
\end{align}
represent the low- and high- frequency components, respectively.
Here, we consider the limit $|g_{s}^{\prime}|/\Delta_{e}\ll1$. By
using a time-averaging treatment~\cite{Xgamel2010time}, the
behavior of $H_{\text{high}}$ can be approximated by a
time-averaged Hamiltonian,
\begin{align}\label{seq:time-averaged_Hamiltonian}
H_{\text{TA}}=\;&\frac{|g_{s}^{\prime}|^2}{2\Delta_{e}}\sum_{k}a_{s}^{\dag}a_{s}\left(|e\rangle_{k}\langle e|-|f\rangle_{k}\langle f|\right)\nonumber\\
&-\frac{|g_{s}^{\prime}|^2}{2\Delta_{e}}\sum_{k,k^{\prime}}\left(|f\rangle_{k}\langle
e|\right)\left(|e\rangle_{k^{\prime}}\langle f|\right).
\end{align}
The first term describes an energy shift depending on the photon
number of the squeezed-cavity mode, and the second term describes
a direct coupling between the two atoms. Accordingly,
$\mathcal{H}\left(t\right)$ becomes
$\mathcal{H}\left(t\right)\simeq H_{\text{low}}+H_{\text{TA}}$,
and after transforming back to the original frame, we obtain
\begin{align}\label{seq:fullHwithtimeaverageeff}
H\left(t\right)\simeq\; &\sum_{k}\left[\Delta_{e}|e\rangle_{k}\langle e|+\Delta_{f}|f\rangle_{k}\langle f|\right]+\omega_{s}a_{s}^{\dag}a_{s}\nonumber\\
\;+\;&g_{s}\sum_{k}\left(a_{s}|e\rangle_{k}\langle f|+\text{H.c.}\right),\nonumber\\
+\;&\frac{1}{2}\Omega_{\text{MW}}\sum_{k}\left(|f\rangle_{k}\langle
g|+\text{H.c.}\right)+V\left(t\right)+H_{\text{TA}}.
\end{align}
We find, from Eq.~(\ref{seq:time-averaged_Hamiltonian}), that the
counter-rotating terms are able to conserve the excitation number
as long as $|g_{s}^{\prime}|/\Delta_{e}\ll1$. Therefore, we can
restrict our discussion in a subspace having at most one
excitation, as discussed above. In this subspace, $H_{\text{TA}}$
is expanded as
\begin{align}\label{SpannedTAH}
H_{\text{TA}}=&-\frac{|g_{s}^{\prime}|^2}{2\Delta_{e}}\left(\mathcal{I}/2+|\varphi_{e}\rangle\langle\varphi_{e}|-|\phi_{+}\rangle\langle\phi_{-}|+\text{H.c.}\right)\nonumber\\
&-\frac{|g_{s}^{\prime}|^2}{\Delta_{e}}\left(\mathcal{I}^{\left(1\right)}/2-|\phi_{+}^{\left(1\right)}\rangle\langle\phi_{-}^{\left(1\right)}|+\text{H.c.}\right),
\end{align}
where
\begin{align}
\mathcal{I}^{\left(1\right)}=\;&|\phi_{+}^{\left(1\right)}\rangle\langle\phi_{+}^{\left(1\right)}|+|\phi_{-}^{\left(1\right)}\rangle\langle\phi_{-}^{\left(1\right)}|
+|\psi_{+}^{\left(1\right)}\rangle\langle\psi_{+}^{\left(1\right)}|+|\psi_{-}^{\left(1\right)}\rangle\langle\psi_{-}^{\left(1\right)}|,\nonumber\\
|\phi_{\pm}^{\left(1\right)}\rangle=\;&\left(|gg\rangle\pm|ff\rangle\right)|1\rangle_{s}/\sqrt{2},\nonumber\\
|\psi_{\pm}^{\left(1\right)}\rangle=\;&\left(|gf\rangle\pm|fg\rangle\right)|1\rangle_{s}/\sqrt{2}.
\end{align}
Equation (\ref{SpannedTAH}) indicates that the counter-rotating
terms introduce an energy shift of
$|g_{s}^{\prime}|^2/\left(2\Delta_{e}\right)$ imposed upon the
ground states, and a coherent coupling, of strength
$|g_{s}^{\prime}|^2/\left(2\Delta_{e}\right)$, between the states
$|\phi_{+}\rangle$ and $|\phi_{-}\rangle$. From
Fig.~\ref{Sfighopping}(a), we find that in the regime, where
$\Omega/|g_{s}^{\prime}|$ is comparable to
$|g_{s}^{\prime}|/\Delta_{e}$, such an energy shift can cause the
$|\psi_{+}\rangle\rightarrow|D\rangle$ transition to become far
off-resonant and, thus, suppress the population into the desired
state $|\psi_{-}\rangle$. Meanwhile, this introduced coupling may
increase the entanglement error originating from the microwave
dressing of the ground states. For example, if
$\Delta_{f}=|g_{s}^{\prime}|^2/\left(2\Delta_{e}\right)$, then the
state $|\phi_{-}\rangle$ becomes a dark state of the microwave
drive. In this case, the population in $|\phi_{-}\rangle$ is
trapped and cannot be transferred to $|\psi_{-}\rangle$. To remove
these detrimental effects, it is essential to compensate this
energy shift. According to the above analysis, the detunings in
Eq.~(\ref{seq:detuningsforeffmeq}) need to be modified as
\begin{align}\label{seq:modification}
\Delta_{e}=\beta-\frac{|g_{s}^{\prime}|^{2}}{2\Delta_{e}}=\omega_{s}+\Delta_{f}-\frac{|g_{s}^{\prime}|^2}{\Delta_{e}}.
\end{align}
This modification simplifies the full dynamics to the same
hopping-like model, as shown in Fig.~\ref{Sfighopping}(a) with
$\Delta_{f}\rightarrow\Delta_{f}^{\prime}=\Delta_{f}-|g_{s}|^{2}/\left(2\Delta_{e}\right)$.
Therefore, we can map the full system to a simple system that
excludes the counter-rotating terms and has been discussed above.

\begin{figure}[tbph]
\centering
\includegraphics[width=17.5cm]{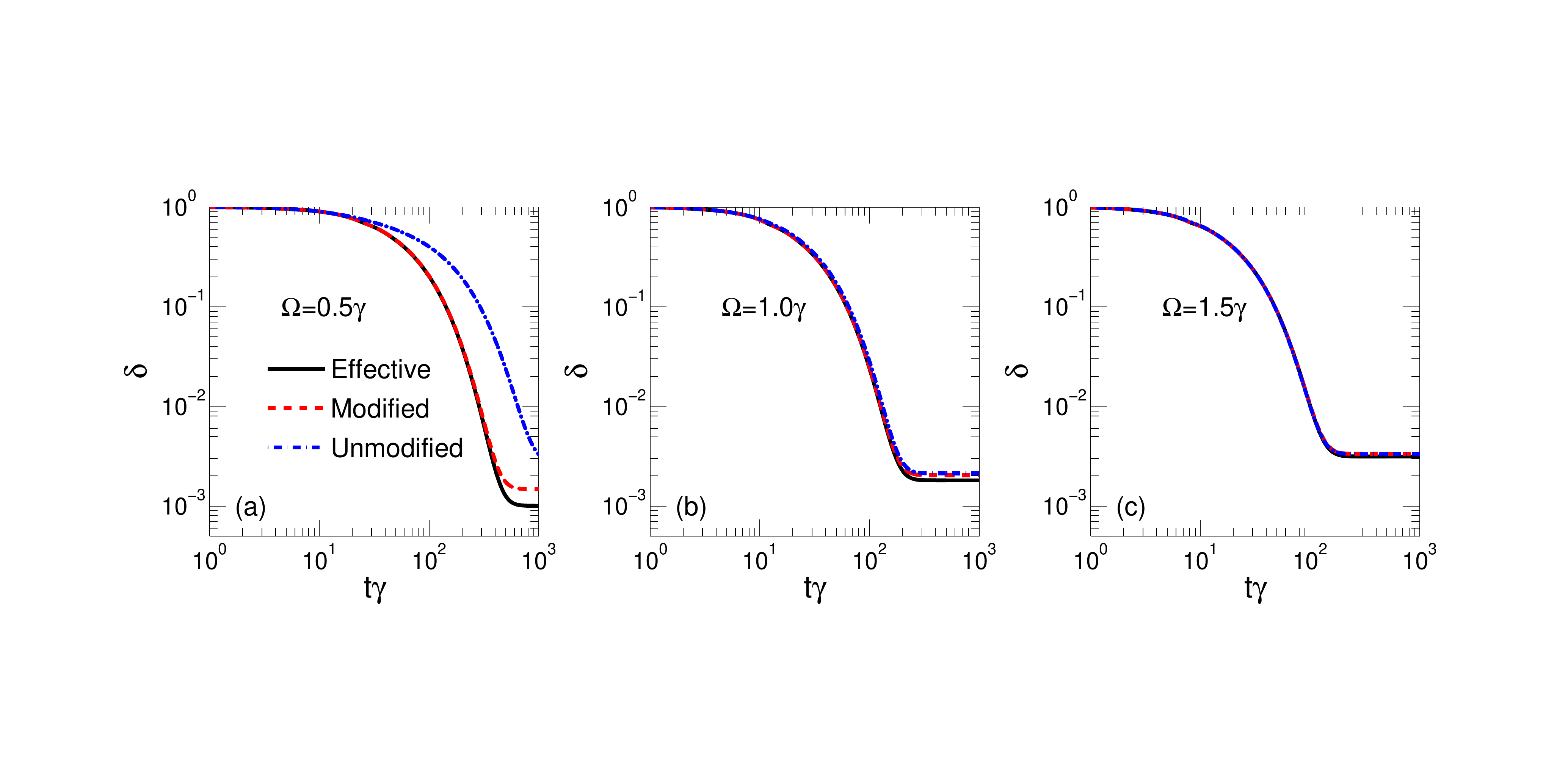}
\caption{(Color online) Entanglement infidelity $\delta$ as a
function of time $t\gamma$ for (a) $\Omega=0.5\gamma$, (b)
$\Omega=1.0\gamma$, and (c) $\Omega=1.5\gamma$, assuming a
cooperativity of $C=20$. Solid and dashed-dotted curves are
obtained, respectively, from integrations of the effective and
full master equations, both with detunings
$\Delta_{f}=\Omega/2^{7/4}$ and
$\Delta_{e}=\beta=\omega_{s}+\Delta_{f}$. Dashed curves are also
given by calculating the full master equation but with modified
detunings
$\Delta_{f}=\Omega/2^{7/4}+|g_{s}^{\prime}|^2/\left(2\Delta_{e}\right)$
and
$\Delta_{e}=\beta-|g_{s}^{\prime}|^{2}/\left(2\Delta_{e}\right)=\omega_{s}+\Delta_{f}-|g_{s}^{\prime}|^{2}/\Delta_{e}$.
For both full cases, we have assumed
$\Delta_{e}=200g_{s}^{\prime}$. In all plots, we have assumed that
$\gamma_{g}=\gamma/2$, $\kappa=2\gamma/3$,
$\Omega_{\text{MW}}=\sqrt{2}\Delta_{f}$, $r_{p}=3$, and
$\theta_{p}=\pi$. Moreover, the initial state of the atoms is
$\left(\mathcal{I}-|\psi_{-}\rangle\langle\psi_{-}|\right)/3$ and
the cavity was initially in the vacuum. }\label{Sfigmodified}
\end{figure}

To understand this process better, we can follow the same method
as above, but now with the Hamiltonian in
Eq.~(\ref{seq:fullHwithtimeaverageeff}). Thus, we find the
effective Hamiltonian and Lindblad operators as follows:
\begin{align}
H_{\text{eff}}^{\prime}=\;&\Delta_{f}^{\prime}\left(\mathcal{I}/2-|\phi_{+}\rangle\langle\phi_{-}|+\text{H.c.}\right)+\Omega_{\text{MW}}\left(|\psi_{+}\rangle\langle \phi_{+}|+\text{H.c.}\right),\\
L_{g1,\text{eff}}^{\prime}=\;&r_{g}^{\prime}\left[\left(|\psi_{+}\rangle+|\psi_{-}\rangle\right)\left(\gamma_{\text{eff},0}^{\prime}\langle \psi_{+}|+\gamma_{\text{eff},2}^{\prime}\langle \psi_{-}|\right)+\gamma_{\text{eff},1}^{\prime}\left(|\phi_{+}\rangle+|\phi_{-}\rangle\right)\left(\langle\phi_{+}+\langle\phi_{-}|\right)\right],\\
L_{g2,\text{eff}}^{\prime}=\;&-r_{g}^{\prime}\left[\left(|\psi_{+}\rangle-|\psi_{-}\rangle\right)\left(\gamma_{\text{eff},0}^{\prime}\langle \psi_{+}|-\gamma_{\text{eff},2}^{\prime}\langle\psi_{-}|\right)+\gamma_{\text{eff},1}^{\prime}\left(|\phi_{+}\rangle+|\phi_{-}\rangle\right)\left(\langle\phi_{+}+\langle\phi_{-}|\right)\right],\\
L_{f1,\text{eff}}^{\prime}=\;&r_{f}^{\prime}\left[\left(|\phi_{+}\rangle-|\phi_{-}\rangle\right)\left(\gamma_{\text{eff},0}^{\prime}\langle \psi_{+}|+\gamma_{\text{eff},2}^{\prime}\langle\psi_{-}|\right)+\gamma_{\text{eff},1}^{\prime}\left(|\psi_{+}\rangle-|\psi_{-}\rangle\right)\left(\langle\phi_{+}|+\langle\phi_{-}|\right)\right],\\
L_{f2,\text{eff}}^{\prime}=\;&-r_{f}^{\prime}\left[\left(|\phi_{+}\rangle-|\phi_{-}\rangle\right)\left(\gamma_{\text{eff},0}^{\prime}\langle \psi_{+}|-\gamma_{\text{eff},2}^{\prime}\langle\psi_{-}|\right)+\gamma_{\text{eff},1}^{\prime}\left(|\psi_{+}\rangle+|\psi_{-}\rangle\right)\left(\langle\phi_{+}|+\langle\phi_{-}|\right)\right],\\
L_{\text{as},\text{eff}}^{\prime}=\;&r_{\text{as}}^{\prime}\left[\kappa_{\text{eff},1}^{\prime}|\psi_{-}\rangle\left(\langle\phi_{+}|+\langle\phi_{-}|\right)-\frac{1}{\sqrt{2}}\kappa_{\text{eff},2}^{\prime}
\left(|\phi_{+}\rangle-|\phi_{-}\rangle\right)\langle
\psi_{-}|\right].
\end{align}
Here,
\begin{align}
\Delta_{f}^{\prime}=\;&\Delta_{f}-\frac{|g_{s}|^{2}}{2\Delta_{e}},\\
r_{g\left(f\right)}^{\prime}=\;&\exp(-i\beta t)\frac{\Omega\sqrt{\gamma_{g\left(f\right)}}}{4\gamma},\\
r_{\text{as}}^{\prime}=\;&\exp(-i\beta
t)\frac{\Omega}{2\sqrt{\gamma}},
\end{align}
and
\begin{align}
\gamma_{\text{eff},0}^{\prime}=\;&\frac{1}{\widetilde{\Delta}_{e}^{\prime}},\\
\gamma_{\text{eff},m}^{\prime}=\;&\frac{\widetilde{\omega}_{s,m}^{\prime}}{\widetilde{\omega}_{s,m}^{\prime}\widetilde{\Delta}_{e,m-1}^{\prime}-mC_{s}},\\
\kappa_{\text{eff},m}^{\prime}=\;&\frac{\sqrt{mC_{s}}}{\widetilde{\omega}_{s,m}^{\prime}\widetilde{\Delta}_{e,m-1}^{\prime}-mC_{s}}
\end{align}
where
\begin{align}
\widetilde{\Delta}_{e}^{\prime}=&\;\left(\Delta_{e}+\Delta_{f}-\beta\right)/\gamma-i/2,\\
\widetilde{\omega}_{s,m}^{\prime}=&\left[\omega_{s}+m\left(\Delta_{f}-\frac{|g_{s}^{\prime}|^2}{\Delta_{e}}\right)-\beta\right]/\kappa-i/2,\\
\widetilde{\Delta}_{e,m-1}^{\prime}=&\left[\Delta_{e}-\beta+\left(m-1\right)\left(\Delta_{f}-\frac{|g_{s}^{\prime}|^{2}}{\Delta_{e}}\right)\right]/\gamma-i/2,
\end{align}
for $m=1,2$. Upon using the modified parameter, given in
Eq.~(\ref{seq:modification}), we obtain
$\widetilde{\Delta}_{e}^{\prime}\sim\widetilde{\omega}_{s,m}^{\prime}\sim\widetilde{\Delta}_{e,m-1}^{\prime}\sim-i/2$.
This implies that the dynamics is the same as what we have already
described for the simplified system without the counter-rotating
terms, thereby leading to the same entanglement infidelity. To
confirm this, we perform numerical calculations, as shown in
Fig.~\ref{Sfigmodified}. Specifically, we plot the entanglement
infidelity as a function of rescaled time. Solid curves indicate
the results obtained by integrating the effective master equation,
whereas dashed and dashed-dotted curves reveal the predictions of
the full master equation, respectively, with modified and
unmodified detunings. These results demonstrate that the
detrimental effects of the counter-rotating terms can be strongly
suppressed by modifying external parameters, in particular, as
what we have discussed above, for the case of weak $\Omega$
driving strengths, which are necessary for the validity of the
perturbative treatment used in our approach.

%

\end{document}